\documentclass[lettersize,journal]{IEEEtran}
\usepackage{amsmath,amsfonts}
\usepackage{array}

\usepackage{textcomp}
\usepackage{stfloats}
\usepackage{url}
\usepackage{verbatim}
\usepackage{graphicx}
\usepackage{cite} 
\usepackage{xspace}
\usepackage[tight,TABTOPCAP]{subfigure}
\usepackage{booktabs}
\usepackage{tabularx}
\usepackage{units}
\usepackage{pifont}

\usepackage[linesnumbered,ruled,vlined]{algorithm2e}
\SetKwInput{KwInput}{input} 
\SetKwInput{KwOutput}{output}

\newcommand{\archname}{\text{ZONIA}\xspace}

\begin{document}

\title{\archname: a Zero-Trust Oracle System for Blockchain IoT Applications}

\author{
    \IEEEauthorblockN{
    Lorenzo Gigli\IEEEauthorrefmark{1},  
    Ivan Zyrianoff\IEEEauthorrefmark{1}, 
    Federico Montori\IEEEauthorrefmark{1},
    Luca Sciullo\IEEEauthorrefmark{1},
    Carlos Kamienski\IEEEauthorrefmark{2},
    Marco Di Felice\IEEEauthorrefmark{1}\\ 
    }
    \IEEEauthorblockA{
    \IEEEauthorrefmark{1} \textit{Department of Computer Science and Engineering, University of Bologna, Italy}\\ 
    \IEEEauthorrefmark{2} \textit{Federal University of ABC (UFABC), Santo André, SP, Brazil}\\
    Correspondent author's email:  lorenzo.gigli@unibo.it
    }
}

\maketitle

\begin{abstract}
The rapid expansion of the Internet of Things (IoT) has led to significant data reliability and system transparency challenges, aggravated by the centralized nature of existing IoT architectures. This centralization often results in siloed data ecosystems, where interoperability issues and opaque data handling practices compromise both the utility and trustworthiness of IoT applications. To address these issues, we introduce \archname (Zero-trust Oracle Network for IoT Applications), a novel blockchain oracle system designed to enhance data integrity and decentralization in IoT environments. Unlike traditional approaches that rely on Trusted Execution Environments and centralized data sources, \archname utilizes a decentralized, zero-trust model that allows for anonymous participation and integrates multiple data sources to ensure fairness and reliability. This paper outlines \archname's architecture, which supports semantic and geospatial queries, details its data reliability mechanisms, and presents a comprehensive evaluation demonstrating its scalability and resilience against data falsification and collusion attacks. 
Both analytical and experimental results demonstrate \archname's scalability, showcasing its feasibility to handle an increasing number of nodes in the system under different system conditions and workloads. 
Furthermore, the implemented reputation mechanism significantly enhances data accuracy, maintaining high reliability even when 40\% of nodes exhibit malicious behavior.
\end{abstract}


\section{Introduction}
 
In recent years, the Internet of Things (IoT) has witnessed exponential growth, fueled by advances in connectivity, computational power, and the miniaturization of technology. This expansion allowed many applications and use cases, from smart cities and industrial automation to smart agriculture and environmental monitoring\cite{tii-2023}. The ability to collect data from diverse sensors and devices has become the key to enabling better decision-making processes across various sectors.

As the IoT ecosystem continues to evolve, so does the complexity and scale of the applications it supports. However, the success of these systems is directly related to the reliability, quality, and transparency of the data being used to make them work. Currently, the IoT landscape is characterized by a high degree of data centralization and opacity. The majority of IoT platforms operate as black boxes, with proprietary protocols and siloed architectures that impede data interoperability and transparency\cite{9199799}. This situation undermines trust in IoT data, particularly in use cases that demand high integrity and verifiability, such as IoT-based smart insurance and environmental monitoring.

To address these challenges, blockchain technology has been proposed as a potential solution\cite{WANG2020100081}. Blockchain, with its decentralized and transparent nature, offers an innovative way to enhance data integrity and trust in the IoT ecosystem. By leveraging blockchain, IoT data can be securely and immutably recorded, thus providing a verifiable and transparent ledger of all transactions and data exchanges\cite{9096382}.

However, integrating blockchain with IoT introduces a unique challenge --- the need for blockchain oracles. Smart contracts, while autonomous, cannot access or interact with data outside their blockchain without help. Oracles serve as bridges between blockchains and the external world, fetching data from off-chain sources to on-chain smart contracts\cite{caldarelli2020understanding}. Yet, the design of a reliable oracle system is far from a trivial challenge\cite{10.1145/3567582}\cite{Sheldon2020Auditing}. At the heart of the oracle problem is the need to trust the oracle service retrieving the data as well as the original data source itself. In addition, for IoT, this trust issue is increased as the oracle system must interact with numerous, heterogeneous, often unreliable sensors and devices\cite{al2022integration}.

In response to these challenges, researchers have proposed various solutions that predominantly rely on Trusted Execution Environments (TEE) or similar technologies to secure the oracle service\cite{provable}\cite{zhang2016town}\cite{chen2021tora}\cite{woo2020distributed}\cite{xian2024distributed}. While these solutions aim to enhance security by providing a protected execution space for code and data, they inadvertently introduce a significant entry barrier for potential new participants wishing to contribute to the system. Moreover, these solutions often depend on single or a limited number of specific data sources for retrieving information. These approaches risk pushing toward a centralization, potentially compromising the system's overall integrity and reliability\cite{9445012}.

Following the vision of our previous work \cite{10230033}, we propose \archname (\textbf{Z}ero-trust \textbf{O}racle \textbf{N}etwork for \textbf{I}oT \textbf{A}pplications), a blockchain oracle system for the retrieval of trusted IoT data. Unlike existing solutions, our approach is based on a decentralized zero-trust paradigm, where individuals can freely join the network without the necessity to reveal their identity or use specialized hardware and be compensated by their data or processing by receiving tokens. Furthermore, \archname is designed to decouple the data sources from the clients, combining results from multiple data sources. This strategy ensures a higher level of decentralization, fairness, and trustworthiness, avoiding single points of failure and enhancing the reliability of the information provided to the clients. 

Our main contributions are threefold:
\begin{itemize}
    \item \textbf{Decentralized Architecture for IoT Data Retrieval}: We introduce the architecture of \archname, which is capable of retrieving IoT data through semantic and geospatial queries.
    \item \textbf{Mechanisms for Data Reliability and System Resiliency}: We detail the implementation of a Truth Inference algorithm and a comprehensive reputation system. These mechanisms work in tandem to evaluate the integrity of data retrieved from distributed IoT sources and to maintain a reliable network of nodes by assessing and updating their reputations based on performance and honesty.
    \item \textbf{Comprehensive System Evaluation}: We present a two-fold evaluation of \archname. First, we assess its performance scalability by examining the system under various configurations and workloads in a real blockchain environment. This demonstrates the system's ability to maintain acceptable performance and scalability, even as the number of participants increases. Second, we evaluate the system's resilience against malicious actors, showcasing its robust mechanisms for ensuring data trustworthiness despite potential collusion or data falsification attempts.
\end{itemize}

The remainder of this paper is structured as follows. Section \ref{sec:architecture} details the architecture of our proposed system, including the roles of Consumers, Producers, Indexers, and Oracles. Section \ref{sec:query-resolution} describes the entire request resolution process, focusing on the node selection and truth inference algorithms and the reputation system. Section \ref{sec:performance} reports on the evaluation of our system in terms of performance and scalability, while Section \ref{sec:security} dives deep into security aspects, assessing the system's capabilities to maintain valuable results even in the presence of malicious actors. Section \ref{sec:related-work} provides a comprehensive review of related work, highlighting the strengths and limitations of existing approaches, and finally Section \ref{sec:conclusions} concludes the paper and outlines directions for future research.

\section{Architecture}\label{sec:architecture}
\begin{figure*}[ht!]
    \centering
    \includegraphics[width=\textwidth]{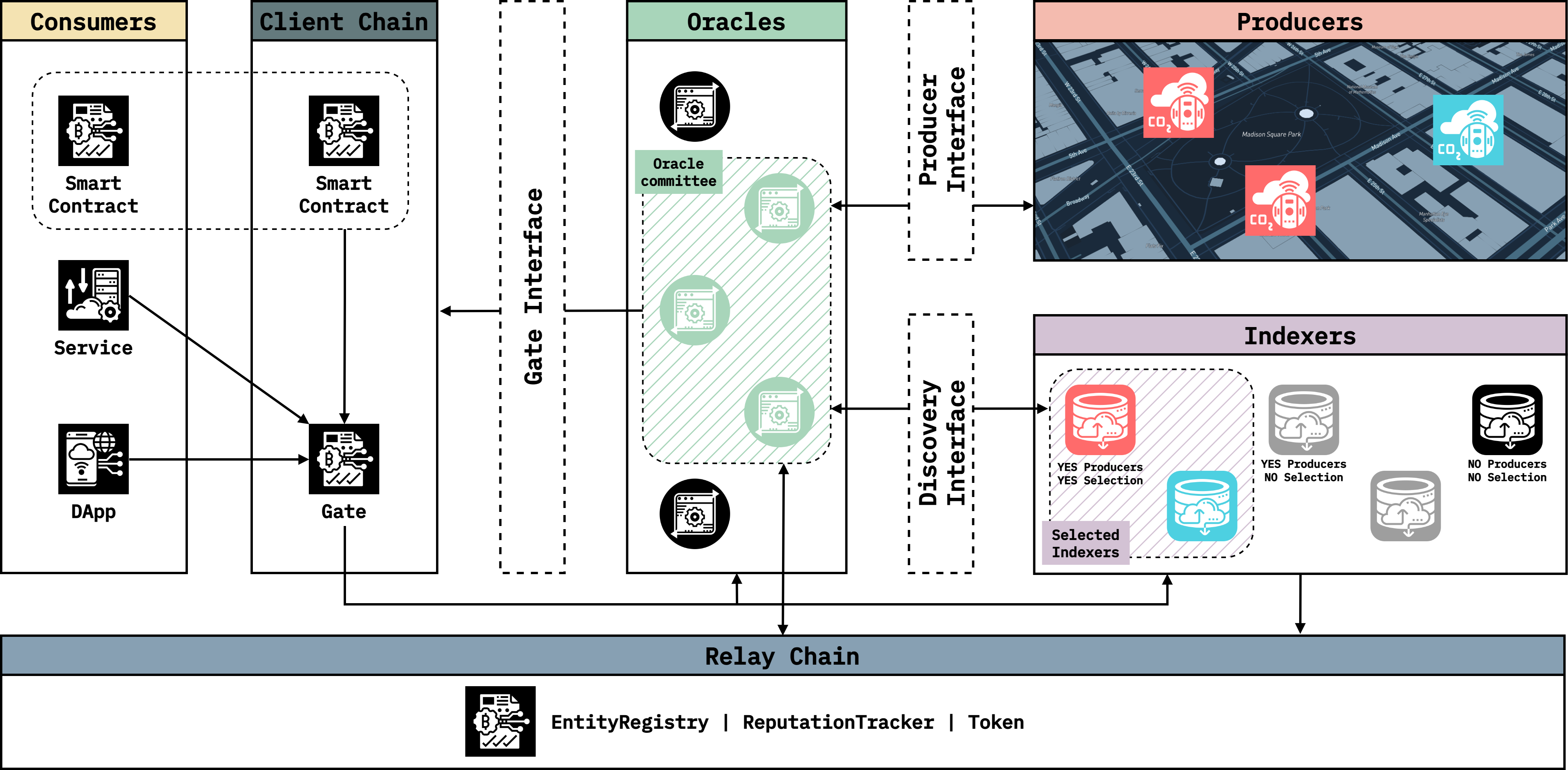}
    \caption{The \archname high level architecture}\label{fig:high-level-architecture}
\end{figure*}

\archname is a distributed oracle system that retrieves data from real-world sensors and devices. Its high-level architecture is illustrated in Fig. \ref{fig:high-level-architecture}, with the following main features:

\begin{itemize}
    \item \textbf{Zero-Trust Decentralized Design}: The system maintains decentralization, allowing all actors to freely join and leave with minimal entry barriers based solely on staking tokens. No specialized hardware, advanced computing capabilities, or complex configurations are required.
    \item \textbf{Semantic and Geospatial Requests}: \archname supports semantic and geospatial requests, enabling precise and context-aware data retrieval. This is critical for applications requiring specific data types and localized information, such as environmental monitoring and automatic insurance.
    \item \textbf{Fair and Secure Node Selection}: The nodes involved in the request resolution are selected using a combination of randomness and reputation scores, ensuring a good balance between unpredictability and fairness in the selection process, mitigating the risk of collusion.
    \item \textbf{Hidden Data Sources}: Data sources within \archname are not directly exposed to clients. Clients cannot choose specific sources for their data needs. Instead, \archname queries multiple data sources that are semantically and geographically compatible with the user's request. An algorithm composes the final result from the collected data, ensuring that the result itself remains decentralized and less susceptible to manipulation.
    \item \textbf{Reputation System}: \archname incorporates a comprehensive reputation system that evaluates and maintains the reliability of network nodes based on their performance and honesty. This system includes anti-tampering measures to prevent manipulation and ensure that only trustworthy nodes participate in data retrieval and request resolution processes.
\end{itemize}

The subsequent sections will delve into the specific actors involved in the system and the integral components that constitute the \archname architecture.

\subsection{Actors}\label{subsec:arch-actors}
\archname has different actors with distinct incentives to keep the system in operational conditions.

\textbf{Consumers} initiate data requests, paying for trustworthy IoT data that are essential to their computational tasks. While smart contracts are the predominant type of Consumer, the category can also include web services, middleware, mobile applications, and even specific IoT devices that require particular data.

\textbf{Producers} are the primary sources of tangible, real-world information processed by the system. They supply the IoT data Consumers seek. They can be static sensors bound to specific locations, dynamic devices like sensor-equipped drones capable of adapting and capturing data from diverse environments, or even real users who join the network through their smartphones and participate in data collection. Producers are rewarded for their contributions to ensure a sustainable ecosystem where quality data provision is consistently incentivized.

\textbf{Indexers} 
keep track and 
maintain metadata about Producers, storing both semantic and geospatial information. They keep this information up to date and decide which Producers to index. It is their responsibility always to provide high-quality Producers to the system. 

\textbf{Oracles} act as intermediaries, processing Consumer requests and interacting with Indexers to identify suitable Producers efficiently. They collect IoT data from these Producers, evaluate it, and execute a consensus process to ensure data integrity. Oracles are rewarded for their contributions, and the validated data is securely written on-chain, ensuring its availability for immediate utilization, thus underpinning \archname's reliability and trustworthiness.

\subsection{Components}\label{subsec:arch-components}
The \archname architecture involves several components that enable the fundamental functionalities of the system. Fig. \ref{fig:high-level-architecture} highlight the main components and interactions among them.

\textbf{Relay Chain} is the primary backbone of \archname. It is an EVM-compatile blockchain hosting the core smart contracts of the system. In specific:
\begin{itemize}
    \item EntityRegistry maintains a record of all Oracles and Indexers registered within the system. It also oversees the staking process, which is integral to the operations delineated in Section \ref{subsec:entities-registration}.
    \item ReputationTracker tracks the reputations of all nodes within the system. This information is utilized in various decision-making processes, such as Oracle and Indexer selection.
    \item Token is the native token of the \archname system. It is used for various purposes, such as paying for requests, incentivizing network participants, and penalizing those who engage in malicious behavior.
    \item Gate is a specialized contract that serves as the entry point for requests on other chains. Each supported chain must have a corresponding Gate contract deployed. Consumers submit requests to the Gate contract on the desired chain, and Oracles pick up these requests for execution.
\end{itemize}

\textbf{Gate Interface} is an interface that all Gate contracts must implement, defining a set of methods that allow Consumers to submit and pay for new requests. Simultaneously, it allows Oracles and Indexers to listen to these requests and write the final result to the client chain.

\textbf{Producer Interface} facilitates the integration of Producers with the \archname system. This interface allows interaction with various devices, regardless of their specific interfaces or network protocols. Through this interface, IoT devices provide semantic descriptors that define their capabilities and data formats. This structured approach enables interactions with Indexers and Oracles, ensuring consistency in data retrieval methods from IoT devices. In our implementation, we leverage the W3C Web of Things (WoT) standard\cite{Toumura:23:WTA} concept of WoT Thing Description (WoT TD)\cite{Korkan:23:WTT} to represent the Producer Interface. The TD allows Producers to provide semantic descriptors defining their capabilities and data formats.

\textbf{Discovery Interface} defines the Indexers' API: how Producers are retrieved within the system. It supports semantic queries to find Producers based on their characteristics, such as the type (e.g., temperature sensors). It incorporates geospatial query capabilities to locate Producers based on their geographical locations (e.g., sensors within a specific city). In our implementation, we utilize the WoT Discovery specification\cite{Cimmino:23:WTD} extended with geospatial query capabilities to support these features.

\subsection{Use Cases}\label{subsec:use-cases}
Following the discussion on the high-level architecture, its key actors, and components, this section explores the practical applications of \archname across various sectors. It highlights how the system facilitates autonomous, reliable, and transparent operations in scenarios that demand multi-party agreements based on data-driven conditions. The utility of \archname is showcased through detailed examples, demonstrating its ability to enhance real-world operations effectively.

\textbf{Smart Insurance} represents a high-level concept where \archname can be effectively applied. One notable application is in agricultural insurance, a sector that traditionally faces challenges with labor-intensive claims processes and susceptibility to fraud. In this context, a smart contract that embodies the insurance policy serves as the primary Consumer. It autonomously initiates payments when predefined environmental conditions, such as specific rainfall or wind speed levels, are met. These parameters are monitored through IoT devices such as weather stations, soil moisture sensors, and drones, alongside data contributions from farmers via smartphones. It is essential to highlight that these data Producers might not be directly affiliated with the farmers or the insurance companies, thus promoting an impartial data collection regime. This framework expedites the claims process, enhancing efficiency and reducing operational costs, but also minimizes the potential for fraudulent claims, strengthening the trust and transparency between insurers and the insured.

\textbf{Environmental Monitoring} can also benefits from the implementation of \archname. In this context, \archname provides a framework capable of supporting governmental initiatives to manage public health and safety more effectively. By integrating data from a wide array of sensors distributed across an urban or rural environment, public institutions can leverage this data to monitor key environmental metrics like air quality, noise pollution, and water quality. Conversely, citizens can also use the system directly to collect data regarding critical situations in the city or their areas and then prepare complaint reports to submit to the public offices. The decentralized nature of \archname ensures that this data is not only accurate but also resistant to tampering, maintaining transparency in public reporting and decision-making processes.

\subsection{Entity registration}\label{subsec:entities-registration}
In the \archname network, the registration process for new nodes is mandatory for their participation in the request resolution process. This registration, applicable to both Oracles and Indexers, is a multi-step procedure designed to ensure the credibility and reliability of these entities. Initially, operators are required to generate a Decentralized Identifier (DID)\cite{Reed:22:DI} and its associated DID Document for their node. This document should include, at a minimum, the node's URL and public key. The URL is the node's API access point, enabling interactions with the other Oracles and Indexers. The public key is used during the cryptographic operations, enabling secure communications and verifying the node's identity in transactional and data exchange processes. The node operator then submits the DID by calling a specific function of the EntityRegistry smart contract.

Furthermore, node operators are obligated to stake a certain amount of tokens as part of the registration. This staking mechanism serves as a security measure, deterring malicious activities by posing risks of token burning or redistribution in case of any inappropriate actions. It also reflects the operator's dedication to the node's continuous functionality and availability.

Notably, Producers do not require a registration process. Given the potentially vast number of Producers, direct on-chain registration is impractical due to its high cost. Instead, Producers can embed their payment addresses in their DID Documents. This approach enables them to receive rewards for their data contributions without on-chain registration, thereby maintaining system efficiency and scalability. This consideration is critical, as the number of Producers could surpass that of other nodes by several magnitudes, posing significant challenges to the system's operational costs if handled traditionally.

\section{Request resolution}\label{sec:query-resolution}
\begin{figure}[t]
    \centering
	\includegraphics[width=0.75\columnwidth]{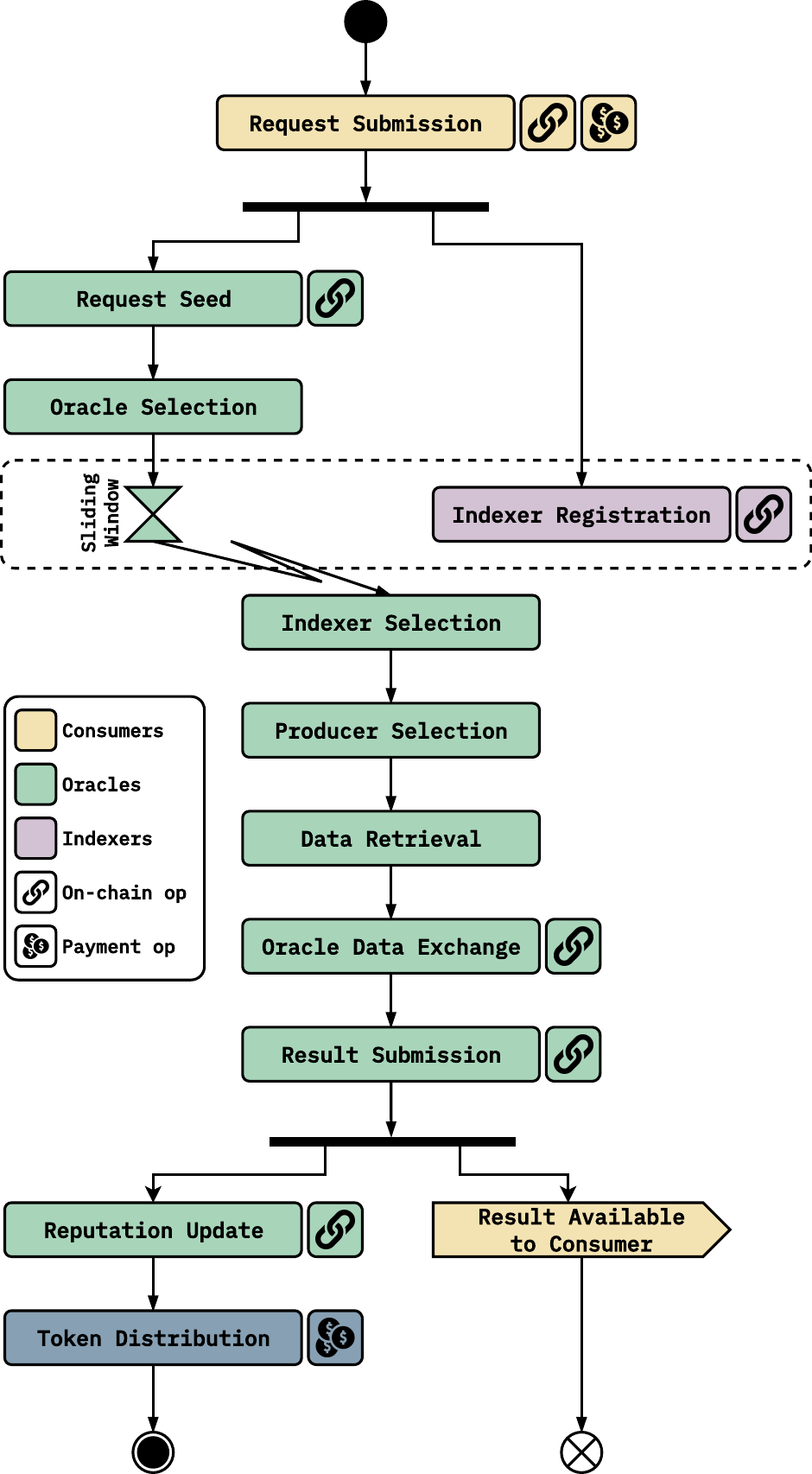}
	\caption{The request resolution flow}\label{fig:query-resolution-flow}
\end{figure}

As illustrated in Fig. \ref{fig:query-resolution-flow}, the request resolution process starts with the \textbf{Request Submission}, where a Consumer submits a request through the Gate smart contract. The request includes a semantic query specifying the data type needed and a geospatial filter to localize the request, for example, to acquire temperature data from the New York City metropolitan area.
After the request submission, Oracle nodes initiate the \textbf{Request Seed} phase, using a competitive process to generate and record a verifiable yet unpredictable seed crucial for subsequent node selection procedures.
Simultaneously, Indexer nodes evaluate their registered Producers to check if some meet the criteria of the Consumer's request. If a match is identified, Indexers proceed with the \textbf{Indexer Registration}, effectively signalling  their capability and intent to participate in the resolution of the request.

After the requested seed, the \textbf{Oracle Selection} occurs, establishing an Oracle committee to manage the submitted request. A sliding time window is initiated, allowing for a grace period during which Indexer nodes may conclude their on-chain registration.
After the expiration of this window, the \textbf{Indexer Selection} takes place, where the committee selects a subset of Indexers from those registered. 
The selection of Oracle and Indexer nodes is strategically based on a combination of randomness and node reputation, ensuring a balance between fairness and trustability. The details of this selection process can be found in Section \ref{subsec:node-selection}.

After that, in the \textbf{Producer Selection}, the committee queries these Indexers to obtain the descriptors of Producers that match the initial request. These descriptor provide structured and comprehensive metadata about the Producers' capabilities and data offerings.
The subsequent \textbf{Data Retrieval} involves the committee querying the Producers for actual sensor data. Following this, during the \textbf{Oracle Data Exchange}, the Oracle nodes of the committee share the gathered data among themselves (Section \ref{subsec:data-gathering}). 
After sharing the data, the committee performs the Truth Inference algorithm, which is needed to evaluate the received data from the Producers and produce a single response. If we take the temperature in New York as an example, the Truth Inference algorithm would output a single temperature value, which would then be the actual reply to the initial request. Detailed descriptions of the Truth Inference algorithm can be found in Section \ref{subsec:truth-inference}.
The process advances to \textbf{Result Submission}, where the Oracle committee writes the agreed-upon result on-chain upon reaching a consensus on the data. Consequently, the result becomes available to the Consumer.

The system then proceeds to the \textbf{Reputation Update}, which adjusts the reputations of all nodes based on their participation in the request resolution. Section \ref{subsec:reputation} discusses additional details on the Reputation system. Finally, \textbf{Token Distribution} occurs, allocating the initial Consumer's payment equitably among the Oracle committee, the participating Indexers, and the Producers. Though not registered on-chain, Producers receive payments directed to their addresses retrieved from the resolved DIDs within the Producers descriptors. The specific payment address for each Producer is located within its DID Document.
This approach facilitates seamless transactions and enhances privacy for the Producers. Being off-chain and having control over their DID Documents, Producers can maintain privacy and have the flexibility to change their payment addresses as needed, further strengthening the security and autonomy within the system.

Furthermore, it is essential to note that this paper does not delve into the specifics of the token distribution among the various actors or the overall economic structure of the system. While these elements are essential to the system's operation, they are outside the scope of this paper's focus, centred on the technological and operational aspects of the request resolution process.

\subsection{Selection Process}\label{subsec:node-selection}
The Selection process is a critical operation of \archname, that takes place for selecting both the Oracle and Indexer nodes participating in the request resolution process (Oracle Selection and Indexer Selection in Fig.~\ref{fig:query-resolution-flow}). Oracles and Indexers are subject to this process differently. The process starts when a Consumer submits a request $\mathbf{R}=\{dtype^{\mathbf{R}}, geo^{\mathbf{R}}, K_{\mathcal{O}}, K_{\mathcal{I}} \}$, where $dtype^{\mathbf{R}}$ is the type of data requested (\textit{e.g.} temperature), $geo^{\mathbf{R}}$ is the geographical area pertaining the request (\textit{e.g.} the New Your city area) and the parameters \( K_{\mathcal{O}} \) and \( K_{\mathcal{I}} \) respectively indicate the number of Oracles and Indexers the Consumer is willing to pay for request resolution. 
The parameters \( K_{\mathcal{O}} \) and \( K_{\mathcal{I}} \) not only determine the scale of the request resolution process, but also directly influence the reliability of the results:
higher \( K_{\mathcal{O}} \) and \( K_{\mathcal{I}} \) values, entailing a higher payment, involve more Oracles and Indexers in the process, enhancing the request resolution's robustness and trustworthiness. This is because a larger pool of nodes contributes to a more comprehensive data aggregation and reduces the impact of any single node's failure, inaccuracy, or malicious behavior.
Conversely, lower \( K_{\mathcal{O}} \) and \( K_{\mathcal{I}} \) values, while more cost-effective, might result in quicker but potentially less reliable results due to the involvement of fewer Oracles and Indexers. The selection of Oracle nodes is an automated process driven by a combination of probability and the reputation of the nodes. This ensures that all Oracle nodes, which are functionally equivalent, have a fair chance of being chosen, with a higher likelihood for those with better reputations. In contrast, the selection of Indexer nodes is more competitive: Indexer nodes, which are responsible for providing access to the relevant IoT data Producers, must first declare their suitability to the request (i.e., that they are indexing enough Producers that return data of type $dtype^{\mathbf{R}}$ within the area $geo^{\mathbf{R}}$), then a selection process akin to the one used for Oracles takes place among these candidates.
\archname is designed to execute these two selection processes in parallel, maximizing the system's efficiency.

Both the selections of Oracles and Indexers involve a probabilistic factor, that is calculated on-chain, must be agreed upon, and should not imply direct communication between nodes, thus it needs to be deterministic.
For this reason, in this phase we make use of Verifiable Random Functions (VRFs) \cite{814584}, which serve as cryptographic primitives that combine the properties of pseudorandom functions and digital signatures. A VRF allows an entity to compute a pseudorandom output $v$ for a given input number $x$ and generate a proof $p$ of its correctness. This way, any entity can verify that $v$ has been deterministically generated with $x$ as input. In the Appendix, we delve into more details about what VRFs are.

\subsubsection{Oracle selection}\label{subsubsec:oracle-selection}
Consider \( \mathcal{O} = \{O_1, O_2, \ldots, O_n\} \) as the set of all Oracle nodes registered to the system. 
Every Oracle node  \( O_i \in \mathcal{O} \) retains a reputation value, defined as \(\vartheta_i \in [-1, 1] \), which indicates how well the Oracle node performed in the past.
Upon the submission of a request, each Oracle receives an event initiating the selection process. The process is non-interactive, deterministic, and fully verifiable, aimed at electing a committee of Oracle nodes \( \widehat{\mathcal{O}} \subseteq \mathcal{O} \) responsible for resolving the specific request \( \mathbf{R} \).

For each node \( O_i \), the Oracle selection process starts by executing a VRF.
As said, in \archname, VRFs are necessary to compute pseudorandom values -- an essential part of the selection process -- within the blockchain.  
With respect to the definition of VRFs, we consider \( x = \mathbf{R}' \), with \(\mathbf{R}' \) defined as the hash of the request \( \mathbf{R} \), to be the given input of the VRF, so that any entity executing it would always obtain the same results. Each Oracle then executes the VRF obtaining \( (v_i, p_i) \), where \( v_i \) is the pseudorandom value and \( p_i \) is the proof of its correctness, both generated by Oracle \( O_i \).
The nodes compete to write their respective VRF output \( (v_i, p_i) \) on the blockchain. The blockchain accepts only the first successfully written VRF output, denoted as \( (v^*, p^*) \), and their submitter receives a sum of tokens as a reward. 
\( v^* \) is then used as the seed for the subsequent steps of the process. This mechanism effectively minimizes on-chain writes; once a winning VRF output is recorded, other nodes cease their submission attempts.

After establishing the seed, each Oracle node first calculates a score \( SO_i \) for every Oracle node using the formula in Equation~\ref{eq:ns-score}:
\begin{equation}\label{eq:ns-score}
    SO_i = \hat{\vartheta_i} \cdot r_i
\end{equation}
where \(\hat{\vartheta_i}\) is the reputation of \(O_i\) normalized between 0 and 1, and \( r_i \in [0,1] \) is a pseudorandom value generated with the winning VRF output \( v^* \) as the seed. 
Finally, each Oracle node then sorts the calculated scores \( SO_i \) in descending order selecting the top \( K_{\mathcal{O}} \) Oracles nodes with the highest scores as the committee for resolving the request. This selection process ensures a balance between the reputation of the nodes and randomness, thereby maintaining fairness and unpredictability in the selection process. The use of a VRF guarantees that all Oracles obtain the same sequence of pseudorandom values, thus they all can independently select the same Oracle committee \( \widehat{\mathcal{O}} \) without interacting with each other.

Based on the above description, Algorithm \ref{algo:oracle-selection} provides a detailed summary of the selection process.

\begin{algorithm}\label{algo:oracle-selection}
\DontPrintSemicolon
\caption{Oracle selection algorithm, executed by each \( O_i \in \mathcal{O} \)}

\KwIn{Oracle nodes $\mathcal{O}$; request $\mathbf{R}$; parameter $K_\mathcal{O}$}
\KwOut{Selection of a committee of $K_\mathcal{O}$ Oracle nodes}

$\mathbf{R}' \gets \text{SHA-256}(\mathbf{R})$\;
$(v_i, p_i) \gets \textsc{VRF}(\mathbf{R}')$\;
Attempt to write $(v_i, p_i)$ to the blockchain\;
$(v^*, p^*) \gets$ First successfully written VRF output on the blockchain\;

\For{$O_i \in \mathcal{O}$}{
    Calculate the normalized reputation $\hat{\vartheta_i}$\;
    Calculate the pseudorandom $r_i$ using $v^*$ as seed\;
    Calculate the score $SO_i$ using Eq. (\ref{eq:ns-score})
}

Sort Oracles in $\mathcal{O}$ based on scores $SO_i$ in descending order\;
Select top $K_\mathcal{O}$ nodes from sorted list\;

\Return{Selected $K_\mathcal{O}$ Oracle nodes \( \widehat{\mathcal{O}} \)}
\end{algorithm}

\subsubsection{Indexer selection}\label{subsubsec:indexer-selection}
Upon the arrival of a request, all Indexer nodes in the network denoted as \( \mathcal{I} = \{I_1, I_2, \ldots, I_m\} \) can read the request details on the blockchain. Similarly to Oracles, every Indexer node  \( I_j \in \mathcal{I} \) retains a reputation value, defined as \(\varrho_j \in [-1, 1] \), needed for performing the Indexer selection. After reading the request, each Indexer node verifies if any of its registered Producers matches with the request constraints $dtype^{\mathbf{R}}, geo^{\mathbf{R}}$.
The system exposes a nominal parameter $M$, which is the number of Producers that each Indexer must return for each request it commits to reply to. 
If an Indexer node \( I_i \) identifies at least $M$ Producers that match the request constraints, it can decide to subscribe on-chain to that specific request. This subscription acts as a declaration of intent to the Oracle committee, signaling that \( I_i \) has enough relevant Producers for the request and wants to participate in the resolution process.
The Oracle committee \( \widehat{\mathcal{O}} \) then waits for the closure of a sliding window, which is dynamically adjusted based on the network conditions and previous request response times. This sliding window is calculated using an exponential moving average (EMA) combined with an adaptive threshold. Specifically, the EMA for the $u$-th request's response time \( T_u \) is a convex combination computed as in Equation~\ref{eq:ns-ema},
\begin{equation}\label{eq:ns-ema}
    EMA_i = \alpha \cdot T_u + (1 - \alpha) \cdot EMA_{u-1}
\end{equation}
where \( \alpha \) is the weighting factor \( (0 < \alpha \leq 1) \), with higher values giving more weight to recent data.
The dynamic time window \( W \) for Indexer subscription is then adjusted as in Equation~\ref{eq:ns-window},
\begin{equation}\label{eq:ns-window}
    \resizebox{.9\hsize}{!}{$W = \max(W_{min}, \min(W_{max}, W_{base} + \phi \cdot (EMA - EMA_{target})))$}
\end{equation}
where \( W_{base} \) represents a baseline window size, \(\phi\) is a scaling factor for the deviation from a target EMA (\(EMA_{target}\)), and \(W_{min}\) and \(W_{max}\) are the minimum and maximum limits for \(W\). This adaptive approach ensures that the time window for Indexer subscription remains responsive to the varying network conditions, balancing the need for timely request resolution with the opportunity for a broad range of Indexers to participate.

Once the sliding window closes, the Oracle committee \( \widehat{\mathcal{O}} \) initiates a selection process identical to the one described in section \ref{subsubsec:oracle-selection}, but for scoring the Indexer nodes and based on the Indexers' reputation. 
Specifically, they score each Indexer according to Equation~\ref{eq:ns-score-indexers}, and after ordering the scores $SI_j$, they pick the first \( K_\mathcal{I}\) winners, defined as the selected Indexers \( \widehat{\mathcal{I}} \subseteq \mathcal{I} \).
\begin{equation}\label{eq:ns-score-indexers}
    SI_j = \hat{\vartheta_j} \cdot r_j
\end{equation}
For a major detail, Algorithm \ref{algo:indexer-selection} provides a detailed summary of the Indexer selection process.

\begin{algorithm}\label{algo:indexer-selection}
\DontPrintSemicolon
\caption{Indexer selection algorithm, executed by each \( O_i \in \widehat{\mathcal{O}} \)}

\KwIn{Indexer nodes $\mathcal{I}$; request $\mathbf{R}$, parameter $K_\mathcal{I}$}
\KwOut{Selection of $K_\mathcal{I}$ Indexer nodes}

\textit{\{ Each Indexer checks if it has at least $M$ Producers that match $dtype^{\mathbf{R}}$ and $geo^{\mathbf{R}}$, if so it registers to the request, becoming part of $\mathcal{I}_{\text{reg}}$ \}}

Wait for the sliding window closure, calculated using Eq. (\ref{eq:ns-ema}) and Eq. (\ref{eq:ns-window})

\For{$I_j \in \mathcal{I_{\text{reg}}}$}{
    Calculate the normalized reputations $\hat{\varrho_j}$\;
    Calculate the pseudorandom $r_j$ using $v^*$ as seed\;
    Calculate the score $SI_j$ using Eq. (\ref{eq:ns-score-indexers})
}

Sort nodes in $\mathcal{I}_{\text{reg}}$ based on scores $SI_j$ in descending order\;
Select top $K_\mathcal{I}$ nodes from sorted list\;

\Return{Selected $K_\mathcal{I}$ Indexer nodes \( \widehat{\mathcal{I}} \) }
\end{algorithm}

\subsection{Data Gathering}\label{subsec:data-gathering}
In this phase, all Oracles in the committee \( \widehat{\mathcal{O}} \) must query all selected Indexers \( \widehat{\mathcal{I}} \) to obtain $M$ Producers from each of them. Then, each \( O_i \in  \widehat{\mathcal{O}} \) must query each of the obtained Producers, gathering a single data point of type $dtype^{\mathbf{R}}$ from each of them. 

Upon querying Indexers to obtain the reference to the Producers, Oracles in the commitee embed a unique request ID within the query. 
This ID plays a crucial role in maintaining the consistency of responses for identical requests. Indexers in \( \widehat{\mathcal{I}} \) leverage this ID to ensure that subsequent queries with the same ID receive the same set of $M$ Producers. This method simplifies the implementation and significantly enhances the Indexers' performance. By utilizing a caching mechanism, once an Indexer node queries its database for a specific request ID, it can efficiently serve cached data for any future requests bearing the same ID. This also ensures that every Oracle in the committee ends up querying the same $P$ selected Producers. Note that $P$ is ideally equal to $K_{\mathcal{I}} \cdot M$, however, in some rare cases, the same Producer can be returned by more than one selected Indexer, thus $P \leq K_{\mathcal{I}} \cdot M$.

After each \( O_i \in \widehat{\mathcal{O}} \) has locally collected the data from the $P$ Producers, it computes a unique hash of the gathered data points. This hash is then written to the Relay Chain. By recording the hash on the Relay Chain before writing the actual data, the Oracles in the committee signal that they have completed their data retrieval without revealing the specifics of the data itself. This preemptive action prevents other Oracles from potentially copying or replicating the data without querying the Producers, thus maintaining the originality and authenticity of the information sourced by each Oracle node.

Once all Oracles in the committee have recorded the hash (i.e., after exactly $K_{\mathcal{O}}$ hashes are available on the Relay Chain), they all write their actual retrieved data onto the Relay Chain as well. By checking the consistency of the previously submitted hash, each other Oracle can verify that the data has not been modified after submitting the hash and was actually collected prior to that.
Note that there seldom may be Oracles in the committee that fail to deliver the hash within a certain amount of time. For this reason, we set a fixed time window after which the data points submitted by Oracles for which the hash is missing are considered null, otherwise we may incur in a starvation problem (see the Appendix for how null values are handled).
Additionally, in certain cases the requested data may surpass a certain size, and saving it to the Relay Chain may be impractical (e.g., multimedia such as pictures and videos instead of simple data points). In such case we can instead store it onto a Distributed File System (DFS) such as IPFS, and only its CID on the relay chain.

\subsection{Truth Inference}\label{subsec:truth-inference}
At this stage, we assume that each \( O_i \in \widehat{\mathcal{O}} \) has now access to the data obtained by all of them.
Specifically, each Oracle in the committee ends up querying the exact same number $P$ of Producers, however, it may obtain different values than other Oracles from the same Producers. This inherently belongs to the nature of IoT: 
Oracle queries in fact take place at slightly different moments in time, thus triggering different sensor readings, causing nondeterminism. 
Nevertheless, because the time interval between sensor readings by multiple Oracles generated by the same request is supposed to be extremely small, we expect the difference between their value to be negligible, assuming that Producers respond consistently to queries. 

Hereafter, we define the full set of $K_\mathcal{O} \cdot P$ data points on the relay chain, relative to a single request \( \mathbf{R} \), to be organized in a matrix $\mathcal{D}$, such that $\mathcal{D}_{i,j}$ is the data point obtained by the $i$-th Oracle in the committee by querying the $j$-th Producer out of $P$. We also assume to be able to, given a Producer, always know which Indexers returned the reference to such a Producer.
The Truth Inference algorithm is a deterministic procedure that every Oracle in the committee must run over $\mathcal{D}$, to infer a single ground truth. If data points are numeric (e.g., temperature or other sensor values), then it is easier to infer a truth value (e.g., mean, median) while it might not be as easy for multimedia or categorical values. In this paper, we do not establish an all-encompassing truth inference algorithm as it might change on top of the needs, however, we detail its properties as follows:
\begin{itemize}
    \item The Truth inference algorithm is a function $T$, which takes in input the data matrix $\mathcal{D}$ and returns a single data point of type $dtype^{\mathbf{R}}$, which is believed to be the closest to the ground truth. The returned value may be part of the input set (e.g., the median) or not (e.g., the mean).
    \item There exists a function $d$ which takes in input two data points and outputs a numerical value. This function represents the notion of distance between two data points.
\end{itemize}
Note that function $T$ is deterministic, therefore any Oracle executing it on $\mathcal{D}$ must obtain the same output. Such an output is then the final reply to the original request, and, after a round of consensus executed among the Oracles in the committee, the final result is written on the calling chain.

\subsection{Reputation}\label{subsec:reputation}
Once the true value has been inferred, the goal of the system is to rate the contribution of both the selected Indexers and Oracles in the committee. 
In order to do so, the first step is to calculate the \textit{score} and the \textit{deviation} of each of the $K_\mathcal{O} \cdot P$ data points of the matrix $\mathcal{D}$, to trace back the validity of the contributions.

We then define the Score matrix $\mathcal{S}$ to be of the same shape as $\mathcal{D}$. 
A single score value $\mathcal{S}_{i,j}$ (relative to a single data point $\mathcal{D}_{i,j}$) is given by $d(\mathcal{D}_{i,j}, T(\mathcal{D}))$, i.e., the distance between the data point $\mathcal{D}_{i,j}$ and the inferred ground truth $T(\mathcal{D})$. 
We similarly define the Deviation matrix $\mathcal{V}$ to be of the same shape as $\mathcal{D}$.
A single deviation value $\mathcal{V}_{i,j}$ (relative to a single data point $\mathcal{D}_{i,j}$) is given by $d(\mathcal{D}_{i,j}, avg[\mathcal{D}_{*,j}])$, which is the distance between the data point $\mathcal{D}_{i,j}$ and the mean of all values obtained from the $j$-th Producer.

Let us also introduce a utility function $c$ called ``the constraint function''.
This function is used to output a rating, which is deemed positive when the input distance or error is small enough with respect to a threshold, negative otherwise. 
$c$ is defined as follows:
\begin{equation}
c(a, \tau) = \frac{2}{1 + (\frac{a} {\tau})^2} - 1, a\geq 0 \wedge \tau > 0 .  
\end{equation}
Such a function takes in input a positive number $a$ and a threshold $\tau$ and returns a result in the interval $(-1, 1]$. It is easy to see that the result assumes a value of $1$ if the input $a$ is $0$, assumes a positive value if $a<\tau$, and asymptotes to $-1$ as $a$ grows large. 

\subsubsection{Rating the selected Indexers}

To rate an Indexer, we must first rate each of the Producers indexed by such Indexer. 
In order to do so, we define the Producer Rating function, which is based on how much the Producer data is (i) accurate and (ii) consistent over time.
More in detail, let us then define the Producer Rating function $\rho$ of the $j$-th Producer as follows:
\begin{equation}\label{eq:rateidx}
    \rho(j) = \beta_{\rho} \frac { \sum_{i=1}^{K_\mathcal{O}} c(S(i,j), \tau_S )  }{ K_\mathcal{O} } + (1 - \beta_{\rho}) \frac { \sum_{i=1}^{K_\mathcal{O}} c(V(i,j), \tau_V ) }{ K_\mathcal{O} }
\end{equation}
The above function is a convex combination -- balanced by a parameter $\beta_{\rho}$ -- of two members:
\begin{itemize}
    \item The first member is the average, over all values obtained from the $j$-th Producer, of the transformed score $c(S(i,j), \tau_S )$, which outputs a positive rating if $S(i,j)$ is smaller than a tolerance $\tau_S$. This term accounts for how much the values obtained from the $j$-th Producer are close to the inferred ground truth.
    \item The second member is the average, over all values obtained from the $j$-th Producer, of the transformed score $c(V(i,j), \tau_V )$, which outputs a positive rating if $V(i,j)$ is smaller than a tolerance $\tau_V$. This term accounts for how much the values obtained from the $j$-th Producer are close to each other (i.e., how much it is consistent). 
\end{itemize}
Every Indexer is considered to be rated with the average of the rating $\rho(j)$, for each Producer $j$ indexed by such Indexer. Formally, we define the rating of an Indexer $k$ through a function $\rho'$ defined as:
$$\rho'(k) = avg [ \rho(j) ] \text{ s.t. } j \in k$$

\subsubsection{Rating the Oracles in the committee}
In order to rate Oracles, we define the Oracle Rating function, which is based on how much Oracles are (i) fast and (ii) consistent with other Oracles querying the same Producers.
More in detail, we define the Oracle Rating function $\theta$ of a single Oracle $O_i$ as follows:
\begin{equation}\label{eq:rateoracles}
    \theta(i) = \beta_{\theta} c(t_i, \tau_t ) + (1 - \beta_{\theta}) \frac { \sum_{j=1}^{m} c(V(i,j), \sigma[V(*, j)] ) }{ m }
\end{equation}
The above function is a convex combination -- balanced by a parameter $\beta_{\theta}$ -- of two members:
\begin{itemize}
    \item The first member is the transformation $c(t_i, \tau_t )$. $t_i$ is the time difference, in milliseconds, between the submission of the hash of $O_i$ and the submission of the hash of the fastest Oracle. If Oracle $O_i$ was the first to submit its hash for the current request, then $t_i = 0$. $\tau_t$ is a value in milliseconds denoting the maximum acceptable delay for Oracles to complete their task efficiently.
    \item The second member is the average, over all values obtained by $O_i$, of the transformed score $c(V(i,j), \sigma[V(*, j)] )$, which outputs a positive rating if $V(i,j)$ is smaller than the standard deviation of the values in the $j$-th column of matrix $V$. This term accounts for how much the values obtained by $O_i$ are in line with the values obtained by all other Oracles in the committee from the same Producers.  
\end{itemize}

Note that, in few cases, Oracles may write null data points on the Relay Chain. We explore these corner cases further in the Appendix.

\subsubsection{Updating the Reputations}\label{sub:reputation}

Once the rating for Indexers and Oracles is calculated for the present request, the system updates the reputation values for both Indexers and Oracles using the outcomes of functions $\rho'$ and $\theta$. 

In particular, recall that $\varrho_k$ is the reputation value of $I_k$, we will then update it with $\delta_\varrho\varrho_k + (1-\delta_\varrho)\rho'_k$, which is a convex combination, tuned by parameter $\delta_\varrho \in (0,1)$, of the previous reputation $\varrho_k$ and the new rating $\rho'_k$. $\delta_\varrho$ tunes how much the new rating will affect the overall reputation of the Indexer. This also ensures that, no matter the value of $\delta_\varrho$, older ratings will always have less impact than newer ones.
Similarly, recall that $\vartheta_i$ is the reputation value of $O_i$, we will then update it with $\delta_\vartheta\vartheta_i + (1-\delta_\vartheta)\theta_i$, with the parameter $\delta_\vartheta \in (0,1)$ tuning the convex combination.

Optionally, if the reputation of an Indexer or an Oracle falls below a certain threshold, then the designated Indexer or Oracle may be banned from the system, as it is labeled as either malicious or strongly unreliable. This threshold is defined as $\omega \in [-1, 1]$ and its behavior is evaluated in Section~\ref{sec:security}.

\section{Performance Evaluation}\label{sec:performance}

In this section, we evaluated \archname to assess its feasibility and scalability. In detail, we aim to answer the following three questions:
(Q1) \emph{What is the effect of request parameters ($ K_\mathcal{O}$ and $ K_\mathcal{I}$) on the system's performance?}
(Q2) \emph{How does the number of Oracles $\mathcal{O}$ and Indexers $\mathcal{I}$ impact the system's performance?}
(Q3) \emph{How do blockchain networks with different performances affect the system's scalability?}

To answer these questions, we designed and deployed an experimental testbed that utilizes an actual blockchain node. The details of this testbed are presented in Section \ref{sub:setup}, where we describe the environmental setup, experiment parameters, and the implementation of actors and components of the system. Subsequently, Subsection \ref{sub:results} outlines the results of our experiments, which address (Q1) and (Q2).
Finally, Subsection \ref{sub:analytical-results} generalizes the performance and scalability of \archname in scenarios where the employed blockchain network has different transaction times, thus addressing the (Q3).

\subsection{Experimental Setup}\label{sub:setup} 

\subsubsection{Testbed Description}
Each 10-minute experiment consists of clients requesting data that triggers the processing steps of the request resolution flow (Fig. \ref{fig:query-resolution-flow}). We collect the time it takes to process each step.
We emulate clients requesting data through a single service that generates random requests and sends them to the blockchain. The inter-arrival times between requests follow an exponential distribution corresponding to a Poisson process.
All system components were deployed as lightweight containers (e.g., Docker\footnote{https://docker.com}) in a LAN setting. We conducted the experiments within a private OpenStack cloud environment. In one virtual machine, we deployed the \archname actors (i.e., Oracles, Produces, Indexers), while in another (deployed in another host machine), we deployed the blockchain node and a logger layer. Each VM was equipped with 16 Intel Xeon Silver 4208 cores and 24 GB of RAM. The host machines and the local network were exclusively dedicated to our experimental setup to prevent disturbances.

To simulate network delays between services communicating with each other, we assigned a location to each node representing a system actor, and a middleware estimated and applied network latency based on the locations of the two communicating nodes. To mimic real-world conditions, we utilized the locations of active bitcoin nodes\footnote{https://bitnodes.io/}, filtering out those using Tor to mask their positions and select only the ones in Europe. Although the system itself is distributed, the requests are geo-located. Consequently, involving nodes for request resolution that are very far apart would not be reasonable. Therefore, we assume localization bounds for selecting nodes, even if those bounds are continental.

We utilized the standard Ethereum Geth\footnote{https://github.com/ethereum/go-ethereum} as the blockchain node. It was configured to implement the Clique \cite{clique} proof of authority consensus protocol, mirroring the setup of the Ethereum Mainnet,  especially regarding the block production interval. In the experiments, we deployed the EntityRegistry and the Gate as EVM-compatible smart contracts within the blockchain node.

\subsubsection{Metrics}
During the experiments, we collected and stored the timestamp of the processing steps of each request resolution, presented in Section \ref{sec:query-resolution} and illustrated by the activity diagram in Fig. \ref{fig:query-resolution-flow}. The recordings of each node were centralized and synchronized in a logging layer (Fluentd\footnote{https://www.fluentd.org/}).

\subsubsection{Factors and Levels}
Table \ref{tab:factors-levels} depicts the factors and levels utilized in the performance evaluation. We conducted experiments by combining different workloads with remaining factors; in those cases, the value of the other parameters was the default one, as denoted in the table.
Due to testbed limitations, parameters larger than those presented in Table \ref{tab:factors-levels} could not be evaluated. In an actual deployment, the workload of computational nodes would be distributed across several blockchain nodes. 
Each experiment was executed for 10 minutes, followed by a 3-minute cool-down period. All experiments were replicated fifteen times, with a calculated confidence interval of 95\%. After each replication, the system was completely reset, with new reputation scores and locations assigned to each node, and the emulated client was initiated using a distinct random seed.

\begin{table}[ht!]
  \centering
  \caption{Factors and Levels}
  \label{tab:factors-levels}
  \begin{tabular}{l|l|l}
    \toprule
    \textbf{Factor} & \textbf{Level} & \textbf{Default} \\
    \midrule
    Number of Oracles & 25, 50 & 50 \\
    Number of Indexers & 25, 50 & 25\\
    Dimension of Oracle Committee ($K_\mathcal{O}$) & 5, 10, 15, 20 & 15 \\
    Dimension of Indexer Committee ($K_{\mathcal{I}}$) & 5, 10, 15 & 5\\
    Workload (requests per second) & 0.2, 0.4, 0.8, 1.6 & - \\
    \bottomrule
  \end{tabular}
\end{table}

\subsubsection{Oracle Implementation}
The Oracle is a NodeJS service that utilizes the \texttt{ethers}\footnote{\url{https://docs.ethers.org/v5/}} library to interact with smart contracts. It subscribes to blockchain events and, upon receiving notifications, performs a specified processing step. For instance, when a new request is submitted, all Oracles receive an event and initiate the request seed operation, while only the Oracles in the committee of a particular request receive notifications from the Oracle data exchange operations. 
In the experiments, the Oracles skipped the step of hashing the data and wrote it directly into the blockchain. Incorporating the hashing step would lead to an additional transaction generated by each Oracle per request. This additional step would cause a delay in the request resolution time, which is comparable to the delay incurred by the current implementation of Oracle data exchange (though slightly lower, given that only the hashes are written in the chain).

\subsubsection{Indexer Implementation}
The Indexer implements a web service that exposes a REST API, aligning with the Discovery Interface. As previously detailed, this API conforms to WoT Discovery specification, facilitating the location of Producers through both semantic and geospatial queries. Each Producer has its own WoT TD which make it compliant with the Producer Interface.
The likelihood of an Indexer storing a Producer matching a given request is determined by the global parameter \textit{data availability threshold}. Each Indexer generates a uniformly distributed random number between 0 and 1. If this value exceeds the defined threshold, the Indexer registers itself in the relay chain for that specific request. To emulate the participation of several Indexers in the system while instantiating a reduced number of them, we set the \textit{data availability threshold} to 0.5. This higher value enables us to instantiate a smaller number of Indexers that effectively represent a larger number. This approach considers that the average number of Indexers participating in each request impacts system performance rather than the sheer quantity of indexing nodes.

Upon startup, we configure the emulated number of Producers (i.e., WoT TDs) stored in each Indexer's database, which varies from 100 to 1,000, 10,000, and 100,000.
The number of Producers indexed by each service follows a heavy-tailed distribution. Most Indexers handle between 100 and 1,000 Producers, while a small subset indexes 100,000 Producers.
We emulated the response time of each Indexer based on the number of Producers it stores. The emulated response times were determined using results from \cite{zion}, which measured the delay of various WoT Indexer implementations, considering the number of stored Producers and the complexity of Producers' descriptions, calculated based on their size in lines.
Each Indexer responded to a query with a random number of TDs, ranging from 1 to 5. Each TD corresponds to a deployed Producer, which returns a given random value when its interface is queried. 

To ensure similarity with real-world scenarios, we distributed the categories of TDs returned in each query to mimic a Pareto distribution, with 80\% consisting of simple TDs, 15\% of medium TDs, and only 5\% of complex TDs. The 
TDs were categorized into different complexity levels based on their number of lines to match the distribution mentioned -- e.g., the 80\% of TDs with fewer lines were classified as simple. The indexed TDs represent real IoT devices, which we collected from repositories of official W3C WoT Events publicly available on Github\footnote{https://github.com/vaimee/tdd-workload-generator/tree/main/src/populate-db/examples-td}.

The sliding time window for Indexer selection was set as a constant value for all experiments. This decision was motivated by the following: (\textit{i}) We aimed to ensure that each experiment represented a portion of time within the system in a normal state. Therefore, after a warm-up period during system startup, the time window was expected to stabilize at a given value with minor alterations; (\textit{ii}) preliminary experiments revealed that employing a dynamic time window mechanism obscured bottlenecks in other system components. As a result, this processing step served as a buffer for the overload of other components.  
In all experiments, the size of the sliding time window was set to \unit{8}{s}. 
Preliminary experiments demonstrated that this duration is reasonable and does not exclude many Indexers.

The reputation of each Oracle and Indexer is randomly defined at startup through a Gaussian distribution with an average of 0.7. 
\begin{figure*}[ht!]
    \centerline{
        \subfigure[$K_\mathcal{O}$ (Number of Oracles in the committee) and workload relationship]{
            \includegraphics[width=0.235\linewidth]{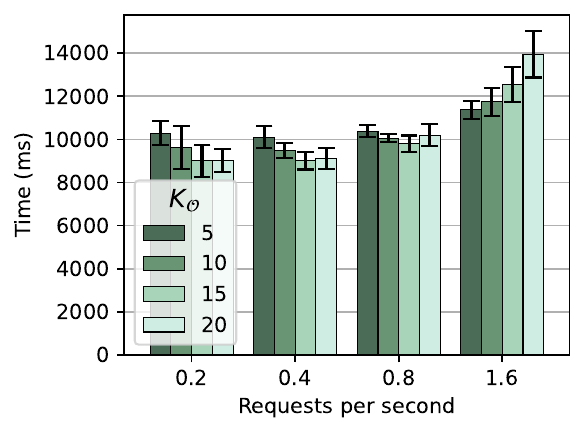}
        }
        \subfigure[$K_{\mathcal{I}}$ (Number of Indexers in the committee) and workload relationship]{
            \includegraphics[width=0.235\linewidth]{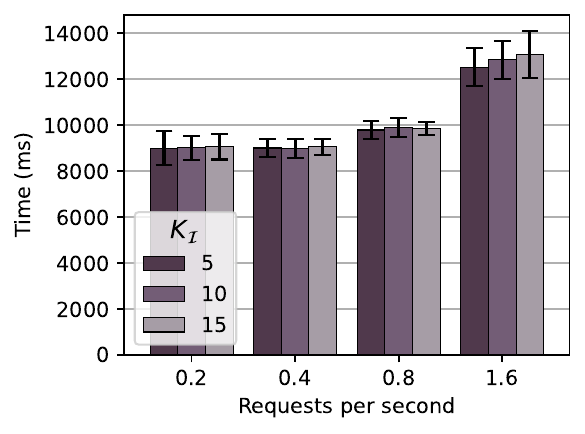}
        }
        \subfigure[Number of Oracles in the system and workload relationship]{
            \includegraphics[width=0.235\linewidth]{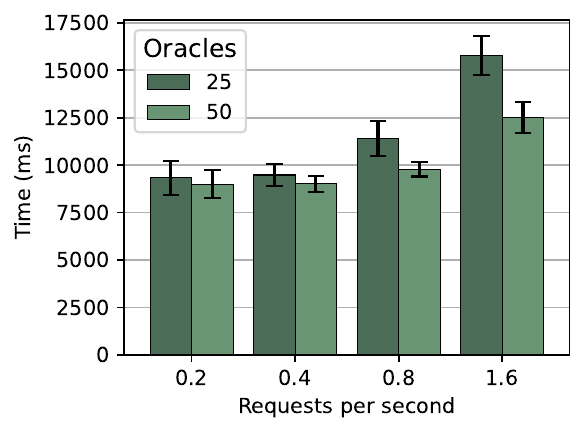}
        }
        \subfigure[Number of Indexers in the system and workload relationship]{
            \includegraphics[width=0.235\linewidth]{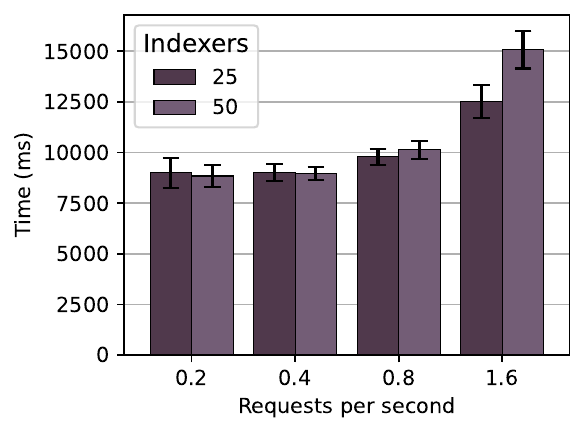}
        }
    }
    \caption{
    Mean end-to-end latency for different parameter combinations.}
    \label{fig:end_to_end}
\end{figure*}

\subsection{Results}\label{sub:results}

\subsubsection{Overview} The end-to-end latency results for each analyzed parameter in an increasing workload are depicted in Fig. \ref{fig:end_to_end} and provide a general overview of the experiment results and the performance of \archname.
The average request resolution time varies from nine seconds under low workloads to fifteen seconds under high workloads. The time is bounded by the relay blockchain transaction throughput of many request resolution steps that generate transactions. 
A blockchain with a higher throughput of transactions per second and lower latency to confirm them, such as Aleph Zero\footnote{https://docs.alephzero.org/aleph-zero/explore/about-aleph-zero}), may significantly impact the processing resolution time and the system scalability. A higher throughput ensures that an increase in workload does not result in a significant increase in latency.
For all parameters, the trend shows an increase in latency for higher workloads, which presents a steep increase as it moves to 1.6 requests per second. This workload is near the system turning point and the limits of the experimental testbed.

Fig. \ref{fig:ko_vs_load}, \ref{fig:ki_vs_load}, \ref{fig:oracle_vs_load}, and \ref{fig:indexer_vs_load} detail the performance of each parameter with varying workloads. In all figures, the top row represents operations that do not generate blockchain transactions, while the bottom row depicts those that do. This distinction is evident when comparing the time scale range on the y-axis of both rows, demonstrating that blockchain operations significantly contribute to the total latency.
Notably, the sum of all processing steps does not add up to the total end-to-end time presented in Fig. \ref{fig:end_to_end}. This discrepancy occurs because some operations are executed parallel or asynchronously, as described in Section \ref{sec:query-resolution}. The experiments reveal that the total request resolution time is primarily determined by the Indexer registration window (constant, defined as 8s) and the operations performed by the Oracles that generate transactions, specifically the Oracle Data Submission and Oracle Final Submission processing steps.

\subsubsection{Size of Oracle Committee versus Workload}
\begin{figure*}[ht!]
	\centering
	\includegraphics[width=0.9\linewidth]{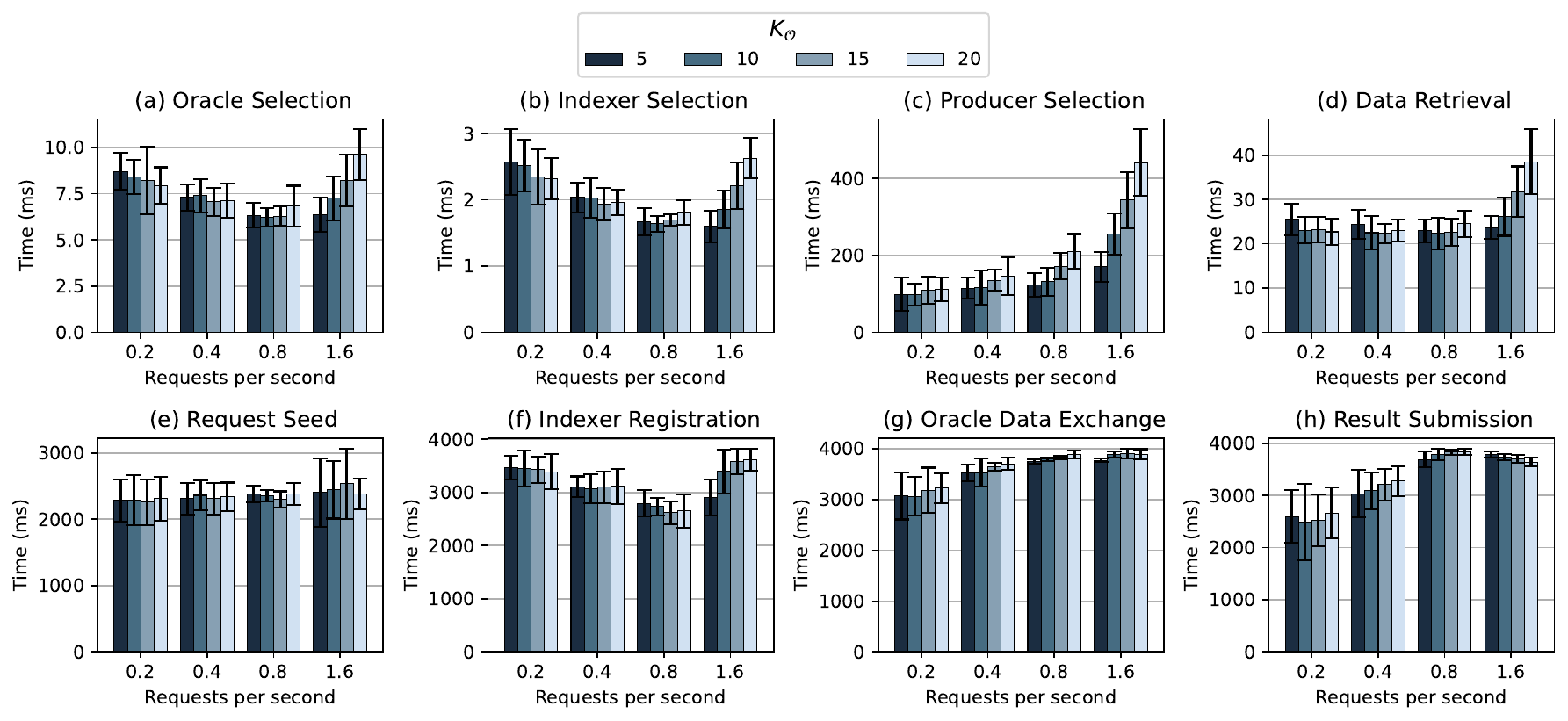}
	\caption{latency for varying  $K_\mathcal{O}$ (Number of Oracles in the committee) versus the workload discrete by processing step.}\label{fig:ko_vs_load}
\end{figure*}

Fig. \ref{fig:end_to_end}a depicts the behavior of different Oracle committee sizes with increasing workload, which is further detailed in Fig. \ref{fig:ko_vs_load}. A larger committee performs better at lower workloads because more transactions are generated, thus filling the blocks faster. However, this trend changes under higher workloads, where the blockchain reaches its maximum block throughput and new transactions are queued. Given the request arrival rate and the chain transaction throughput, this behavior indicates that the system performance can be improved by balancing the number of Oracles in the committee with the workload.

\subsubsection{Size of Indexer Committee versus Workload}
\begin{figure*}[h]
	\centering
	\includegraphics[width=0.9\linewidth]{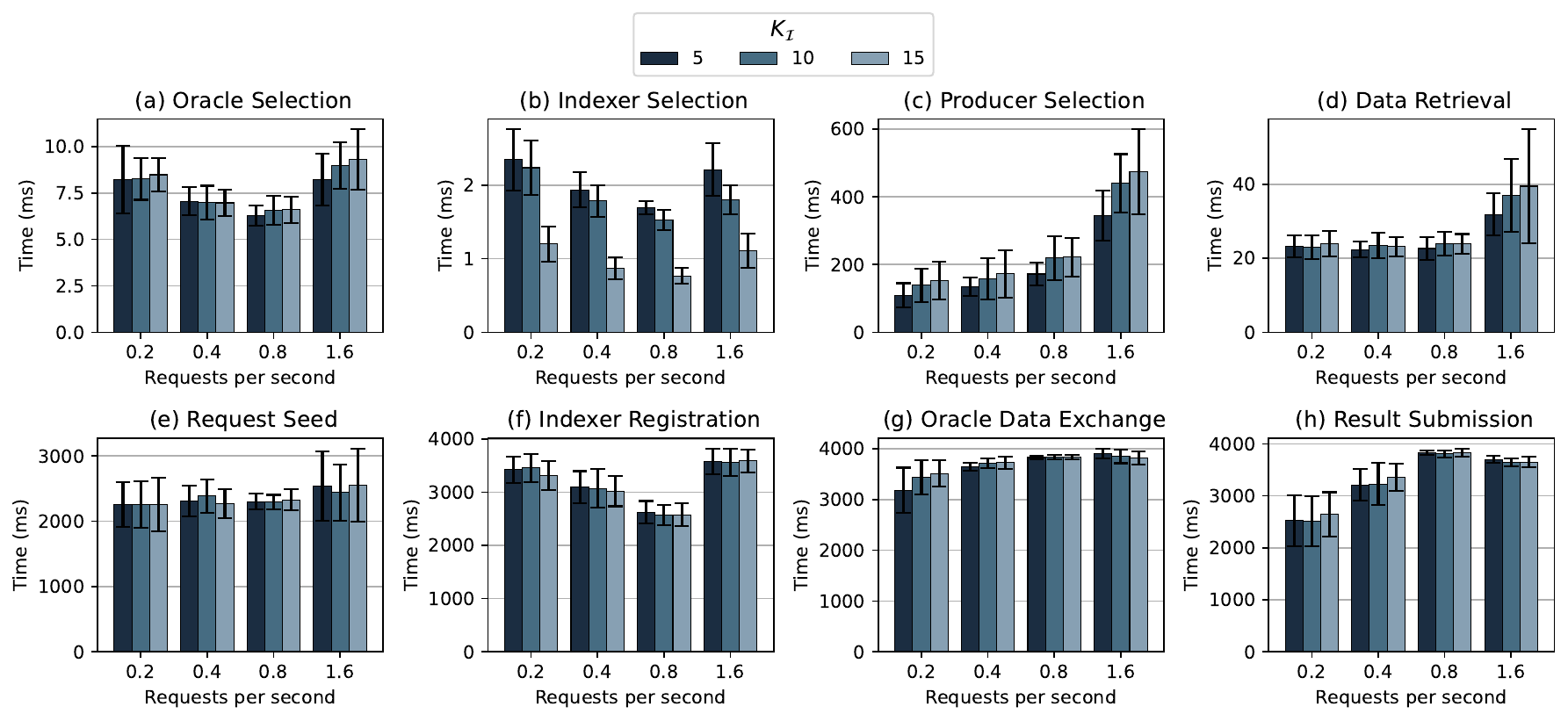}
	\caption{latency for varying $K_{\mathcal{I}}$ (Number of Indexers in the committee) versus the workload discrete by processing step.}\label{fig:ki_vs_load}
\end{figure*}

Fig. \ref{fig:ki_vs_load} corroborates with Fig. \ref{fig:end_to_end}b to evidence that different values of $K_{\mathcal{I}}$ 
do not affect the system performance, which is only impacted by the increase in the workload. Thus, this parameter can be determined based on the client's needs and trustworthiness considerations rather than performance considerations. 


\subsubsection{Number of Oracles versus Workload}

\begin{figure*}[h]
	\centering
	\includegraphics[width=0.9\linewidth]{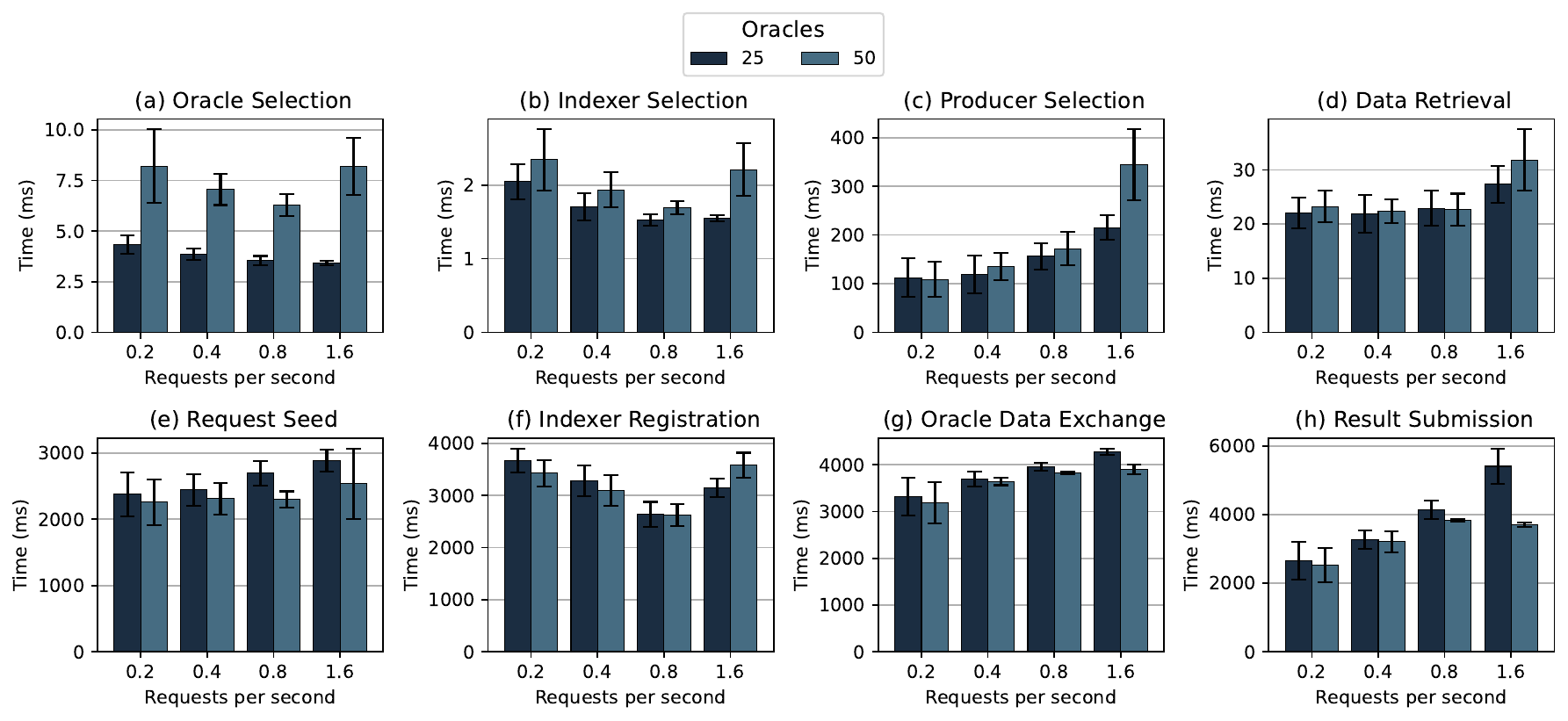}
	\caption{Latency for varying the size of the Oracle population versus the workload discrete by processing step.}\label{fig:oracle_vs_load}
\end{figure*}

Fig. \ref{fig:end_to_end}c illustrates that increasing the number of Oracles positively impacts request latency under higher workloads. When there are more Oracles, the whole request resolution speeds up because the speed of the fastest Oracle bounds two operations: the initial VRF seed and the process of writing true value inferred onto the blockchain -- the last operation is only performed by members of the Oracle committee. This assumption can be confirmed by analyzing Fig. \ref{fig:oracle_vs_load}e, which depicts a lower value when increasing the Oracles, and Fig. \ref{fig:oracle_vs_load}h, which showcases a \unit{2}{s} decrease in the Oracle's final submission. Fig. \ref{fig:oracle_vs_load}a denotes that a higher number of Oracles increases the time to form the Oracle committee. However, this increase is insignificant (less than \unit{10}{ms}), especially when compared to the improvement of the Oracle's final submission.

\subsubsection{Number of Indexers versus Workload}

\begin{figure*}[h]
	\centering
	\includegraphics[width=0.9\linewidth]{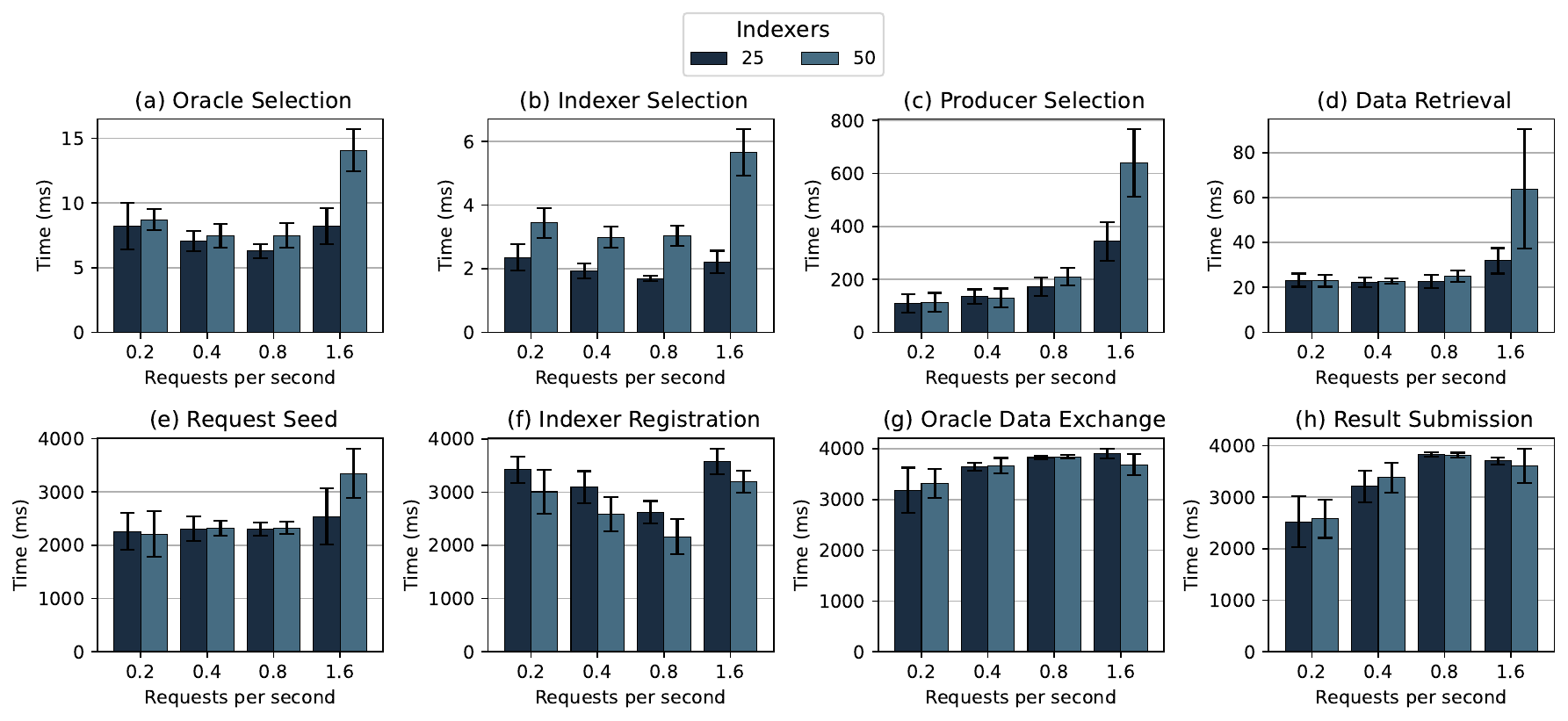}
	\caption{Latency for varying the size of the Indexer population versus the workload discrete by processing step.}\label{fig:indexer_vs_load}
\end{figure*}

In contrast, increasing the number of Indexers in the system harms its performance under higher workloads, as depicted by Fig. \ref{fig:end_to_end}d. In detail, Fig. \ref{fig:indexer_vs_load} demonstrates that the increase in the number of Indexers has a more significant impact on the off-chain operations (i.e., the first row of graphs) than the ones that generate transactions (i.e., the second row of graphs). It is worth noting that the increase in the Indexer population, as modeled in the experiments, affects proportionally the average number of Indexers participating in each request.
This behavior occurs because the increased number of Indexers impacts the average number of Indexers that register to each request since the data availability threshold remains constant at 0.5 in all experiments. We chose to maintain the availability threshold unchanged when increasing Indexers because if the average number of Indexers participating in the request resolution remains constant, increasing the number of Indexers in the system would not impact system performance.

\subsection{Analytical Results}\label{sub:analytical-results}

We further investigate the scalability of the \archname system through an analytical queuing model study. There are two main motivations for this study: (\textit{i}) \emph{Testbed limitations}: The results of our experiments are constrained by the performance of the employed blockchain node, which processes transactions for all system nodes. The workload would be distributed across many blockchain nodes globally in a real blockchain environment. Additionally, the Geth client is known for its slow transaction times \cite{8342866}. (\textit{ii}) \emph{Blockchain heterogeneity}:  there are many different blockchain implementations, and their parameters significantly influence \archname's performance. While our experiments mimic the configuration of Ethereum, Layer-2 blockchains can substantially improve blockchain performance \cite{GANGWAL2023103539}.

We aimed to understand the effect of a blockchain with a faster transaction time on the overall system performance by modelling \archname as a single queue where the service time is the sum of the Oracle processing times and the chain transaction times. Specifically, we consider that the total queuing waiting time $W$ (our end-to-end delay) is given by $W = Wp + Wc$, where $Wp$ is the total processing time of operations that do not interact with the chain and $Wc$ is the total time of operations that generate transactions. In Fig. \ref{fig:ki_vs_load} to Fig. \ref{fig:indexer_vs_load}, the processing time is the sum of Oracle selection, Indexer selection, Producer selection, and data retrieval, whereas chain time is the sum of request seed, Indexer registration, Oracle data exchange, and result submission. Thus, we aim to generalize the results obtained in our real testbed to answer the question (Q3) \emph{How do blockchain networks with different performances affect the system's scalability?}

In the M/M/1 queuing model, $\lambda$ represents the arrival rate (i.e., workload) and, $\mu$ represents the service rate. The waiting time $W$ for an M/M/1 queue is given by Eq. \ref{eq:waiting-time}. In other words, to obtain analytical results for the end-to-end delay ($W$) increasing the workload, we need corresponding service rates $\mu$ for each projected value of $\lambda$.

\begin{equation}\label{eq:waiting-time}
    W = (1/\mu) / (1 – \lambda/\mu)
\end{equation}

We followed a 4-step approach:
\begin{enumerate}
    \item Define values for chain operations $Wc$ assuming faster chains and assuming the processing time $Wp$ coming from our experiments.
    \item Compute values of service rates $\mu$ using the results of our experiments, where we have the workload $\lambda$ and the delay $W$, by isolating $\mu$ in Eq. \ref{eq:waiting-time} and obtaining Eq. \ref{eq:service-rate}.
    \item Forecast $\mu$ values for $\lambda$ values from 0.1 to 10 using regression.
    \item Compute end-to-end delay values ($W$) for the projected $\lambda$ values using Eq. \ref{eq:waiting-time}.
\end{enumerate}

\begin{equation}\label{eq:service-rate}
    \mu = (1 + W * \lambda) / W
\end{equation}

As a deliberate abstraction, we assume that the chain transaction times are constant no matter the workload since the blockchain network should be composed of many nodes distributed across the globe.
We tested individual chain transaction times of 2000, 1000, 500, and 250 ms and used the processing times $Wp$ for the default configurations for Oracles and Indexers given by Table \ref{tab:factors-levels}. Since four operations produce transactions in the critical path for each request to be completed (Fig. \ref{fig:query-resolution-flow}), the total $Wc$ times are 8000, 4000, 2000, and 1000 ms. In most experiments, we calculated the total end-to-end delay $W$ as the sum of those values and the observed processing times $Wp$ for each workload $\lambda$. 
Therefore, we compared six cases: E0, P1, P2, P3, P4. Notice that all represent simulations of our queuing models, but E0 considers $W$ from the experiments, and P1 to P4 assume faster chain transaction times.
\begin{itemize}
    \item E0: end-to-end delay $W$ observed in our experiments
    \item P1: projected value for $Wc$ of 8000 ms plus $Wp$
    \item P2: projected value for $Wc$ of 4000 ms plus $Wp$
    \item P3: projected value for $Wc$ of 2000 ms plus $Wp$
    \item P4: projected value for $Wc$ of 1000 ms plus $Wp$
\end{itemize}

\begin{figure}[h!]
	\centering
	\includegraphics[width=0.75\linewidth]{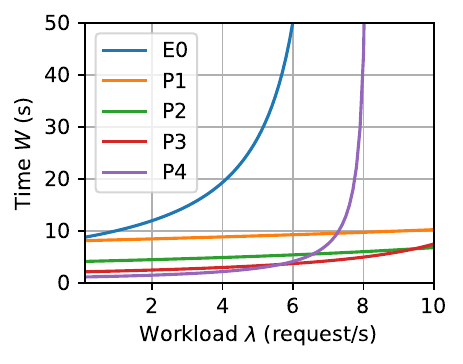}
	\caption{Queuing results for \archname}\label{fig:queue-results}
\end{figure}

Fig. \ref{fig:queue-results} shows the analytical queuing results for the \archname scalability study. 
For the experimental results E0, we can observe that reasonable levels of $W$ can only be obtained for values of $\lambda$ around 4. After that, service rates $\mu$ are smaller than workloads $\lambda$, and the queue grows infinitely so that the system renders unfeasible. On the other hand, for P1 ($Wc=8000ms$) and P2 ($Wc=4000ms$), the time $W$ increases linearly, revealing that it can scale up to 6 times more than what has been tested in the experiments. For P3 ($Wc=2000ms$), we observe that for most workloads $\lambda$, the time $W$ is bounded below 5000ms but starts to move upwards when it reaches 10. Also, a significant result is revealed by P4 ($Wc=1000ms$), where the queue grows indefinitely for workloads $\lambda$ higher than 7, indicating that the bottleneck in this case is the processing time $Wp$ and not anymore the chaining time $Wc$.

These results reveal that the system scalability depends on a balance between processing and chain transaction times. For the chain time $Wc$ observed in the experiments, that system cannot scale up to workload $\lambda=10$. If we consider $Wp=8000ms$, the system scales with higher time $W$, and decreasing $Wc$ yields smaller times, up to a point where the processing time $Wp$ becomes the bottleneck. \textbf{These results lead us to conclude that the higher end-to-end delays of our experimental results are due to the implementation}. The conceptual proposal of the \archname system can solve its main challenge and, at the same time, has room to improve its performance and scalability.
 
\section{Resilience Evaluation}\label{sec:security}
In this section, we assess the capabilities of the system to produce trustworthy results in the presence of malicious actors. The system can be prone to collusion attacks and, as we motivate our work on top of trustworthiness, we need to make sure to punish or kick out actors that misbehave or try to inject false data in the blockchain. In particular, we test our system against a certain number of malicious Producers, Indexers, or Oracles.

\subsection{Experimental Setup}\label{sub:setup-sec}
Differently from the setup shown in Section~\ref{sub:setup}, here we do not need to emulate the real behavior of every single actor, thus we run simulations to represent thousands of actors in a single scenario. 
In particular, we simulate our collusion attack on subsequent client requests for temperature. Let us imagine an example agricultural insurance scenario where the temperature is a determinant factor and must be kept above \unit{20}{\textdegree C}. In our simulation, the current ground truth is set to \unit{25}{\textdegree C}, and the response from the Truth Inference algorithm is considered \textit{accurate} if it falls within a tolerance $\tau_S = $ \unit{3}{\textdegree C} (this is set by the system). A Producer generates its measurements according to a normal distribution with the mean equal to the ground truth, and the standard deviation uniformly generated between 0 and $\frac{4}{3}\tau_S$, representing few honest Producers making occasional mistakes. We also set $\tau_V = $ \unit{1.5}{\textdegree C} to simulate a few honest Producers that will score less due to their lower consistency over time (i.e., higher standard deviation).
We then modeled malicious Producers by setting their mean to a different value, distant from the ground truth, in our case \unit{10}{\textdegree C}, simulating a collusion attack where all malicious actors concur into redirecting the response to a common false value. 

We ran simulations with a fixed population of 100 Indexers at startup.
We randomly sample a subset of them to be malicious, according to the simulation parameter ``Ratio of Malicious Indexers ($RMI$)''. Even though $RMI\in [0,1]$, we never set it above 0.5, as it would automatically compromise the network, according to the well-known 51\% problem \cite{saad2020exploring}, and would not be meaningful for evaluation. For simplicity and without loss of generality, we also set the ratio of malicious Producers to be equal to $RMI$.
Similarly, we also consider the ``Ratio of Malicious Oracles ($RMO$)'' as another simulation parameter.
The number of Oracles in our simulations is set to 20 and, for each request, $K_{\mathcal{O}}=5$ Oracles are selected to reply. Malicious Oracles are randomly sampled at startup and, over the simulation, they try to mechanically tamper the value that they obtain from honest Producers by setting them to a value close to the false truth (\unit{10}{\textdegree C}).

The simulation takes place over 3,000 epochs, each of which corresponds to an entire request resolution process, as per Section~\ref{sec:query-resolution}, outputting a single value. Our simulations feature a population of 500 honest Producers spawned at system startup, which represent a ``warm start''. For the first 1,000 epochs, they build up a positive reputation; this is a realistic assumption for most blockchain-based systems, in fact, they typically run a genesis phase where the blockchain is populated with a solid history of valid blocks in a controlled way. In the remaining 1,000 epochs, we proceed to spawn another 500 Producers, which contain all of the malicious ones, according to a uniform distribution over the epochs. To simulate worst-case collusion attacks, we also perform simulations assuming that all malicious Producers are spawned at the same time. We call this distribution \textit{bursty}, as opposed to the other, which is called \textit{uniform}. Upon its creation, a Producer is assigned to a number of Indexers according to a Pareto distribution with a shape value equal to 2. The Indexers are picked randomly. This is to simulate a high number of Producers that register to only one Indexer and, occasionally, very few Producers that register to multiple Indexers. To simplify our setup, as well as to simulate a worst-case scenario, we assumed that honest Producers only register to honest Indexers, while malicious Producers only register to malicious Indexers. 

Our solution is evaluated against a baseline method, which we call \textbf{\textit{Med}} (referring to ``median''): a deployment of our system where the response to a request is still determined by the truth inference algorithm, but no rating is assigned to either Oracles or Indexers.  This means that every actor has the same chance of being selected in every round, no matter how it behaved in the past. This is compared to our proposed reputation method, which we call \textbf{\textit{Med-R}} (R stands for reputation). In our Med-R we set all values that balance convex combination as fixed, in particular $\delta_\varrho = 0.5$, $\delta_\vartheta = 0.5$, $\beta_\rho=0.9$, and $\beta_\theta=0.25$. Note that this configuration tends to promote how much registered values are coherent with the inferred truth rather than their consistency over time, which is more important when evaluating collusion attacks.
Finally, we evaluate a variant of our proposed method by including the phase where untrusted Oracles and Indexers are banned, according to Section~\ref{sub:reputation}. This method is defined as \textbf{\textit{Med-RB}} (B stands for ban), for which we evaluate different values of $\omega$ (the reputation threshold below which Oracles and Indexers are banned) as a parameter of our simulation.

\begin{figure}
    \centering
    \includegraphics[width=\columnwidth]{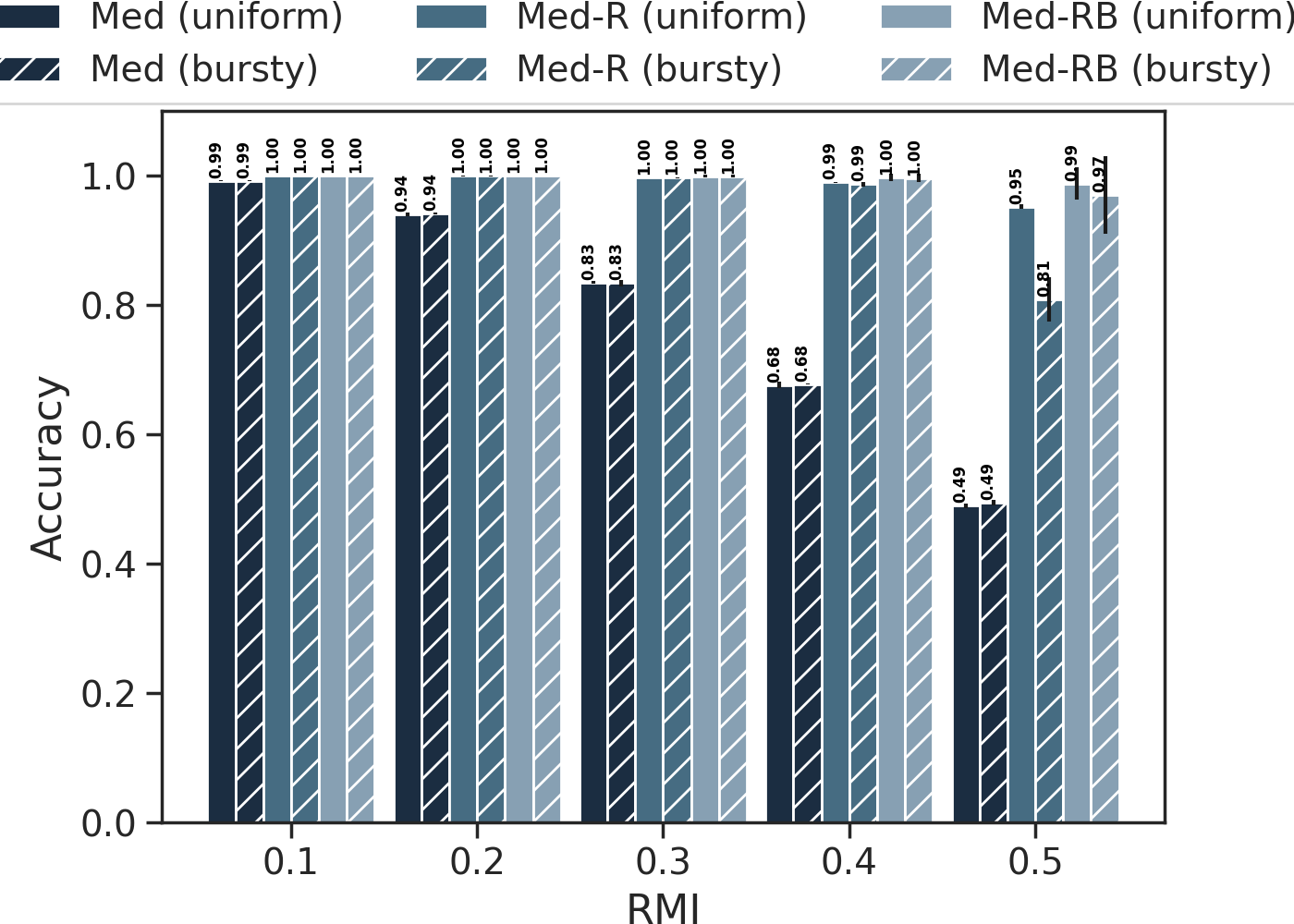}
    \caption{Truth Inference Accuracy: the ratio of client requests that get satisfied within a threshold $\tau_S$ from the ground truth within the last 1,000 epochs.}
    \label{fig:accuracy}
\end{figure}

\begin{figure}[ht]
    \centering
            \includegraphics[width=\columnwidth]{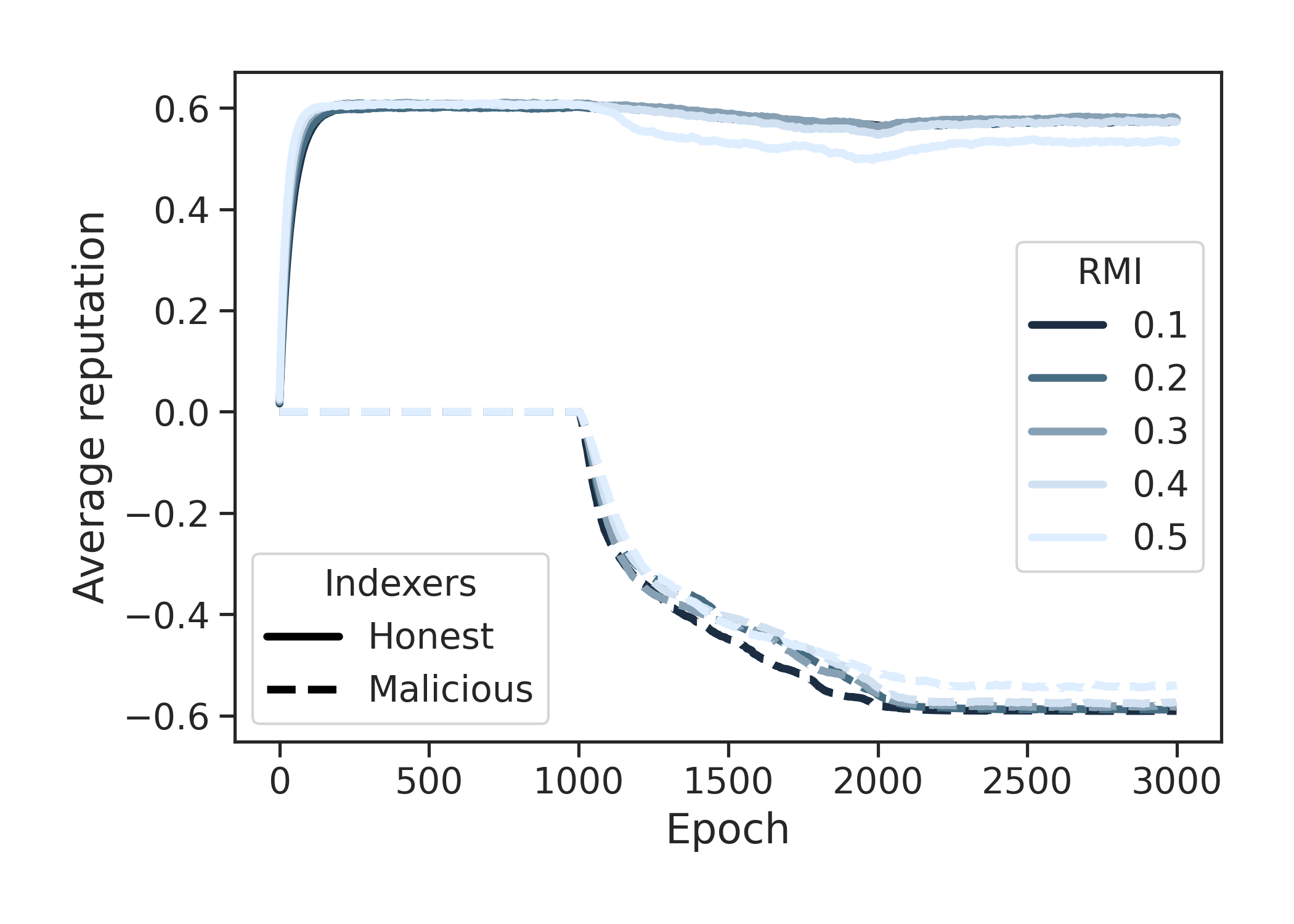}
            \label{fig:rmi-rep}
    \caption{Reputation over time of honest and malicious Indexers over $RMI$}
    \label{fig:blacklisting_RMI}
\end{figure}

\begin{figure*}[hb!]
    \centering
         \subfigure[Blacklisting Recall over $\omega$]{
            \includegraphics[width=0.23\textwidth]{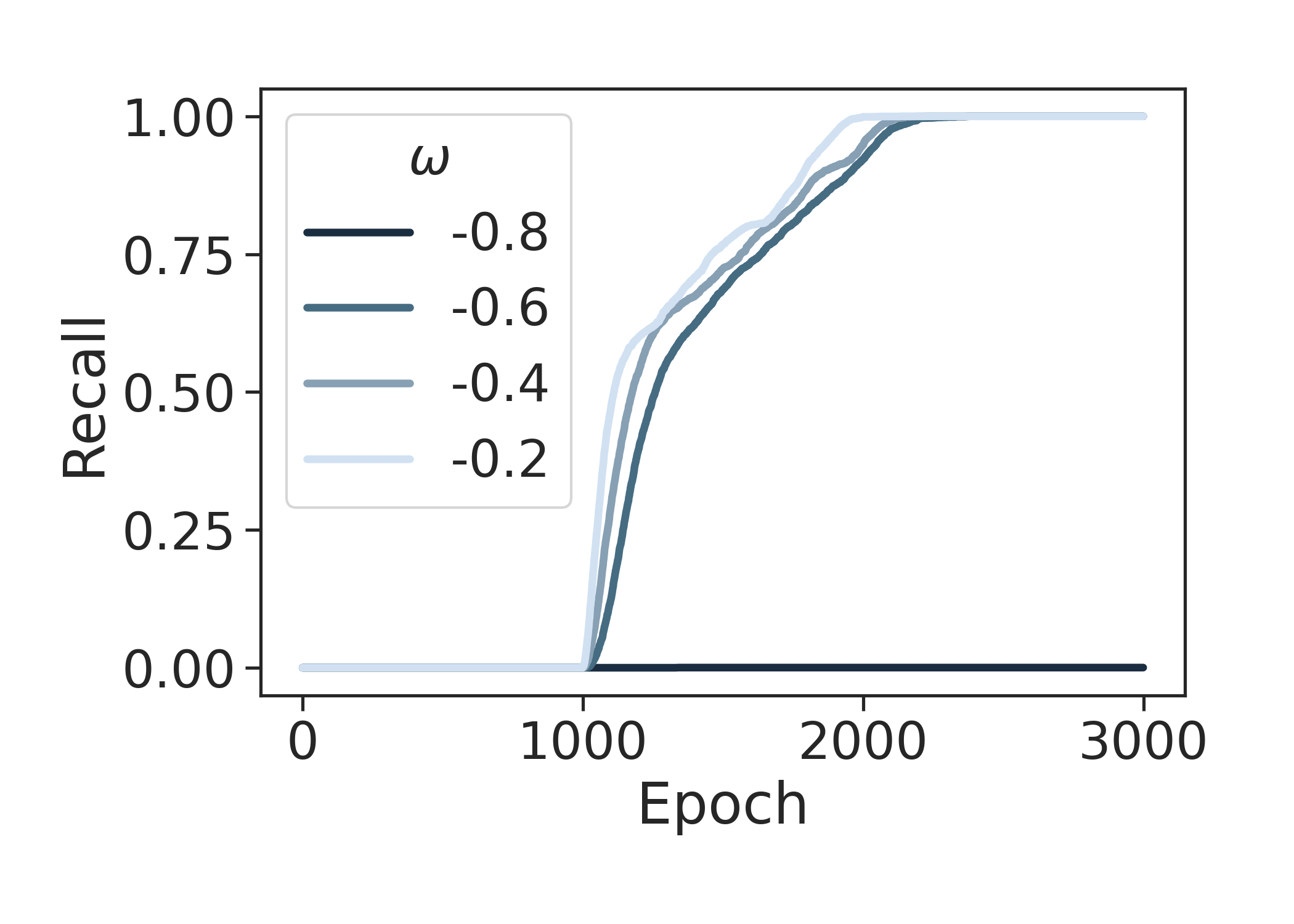}
            \label{fig:omega-rec}
        }
         \subfigure[Blacklisting Precision over $\omega$]{
            \includegraphics[width=0.23\textwidth]{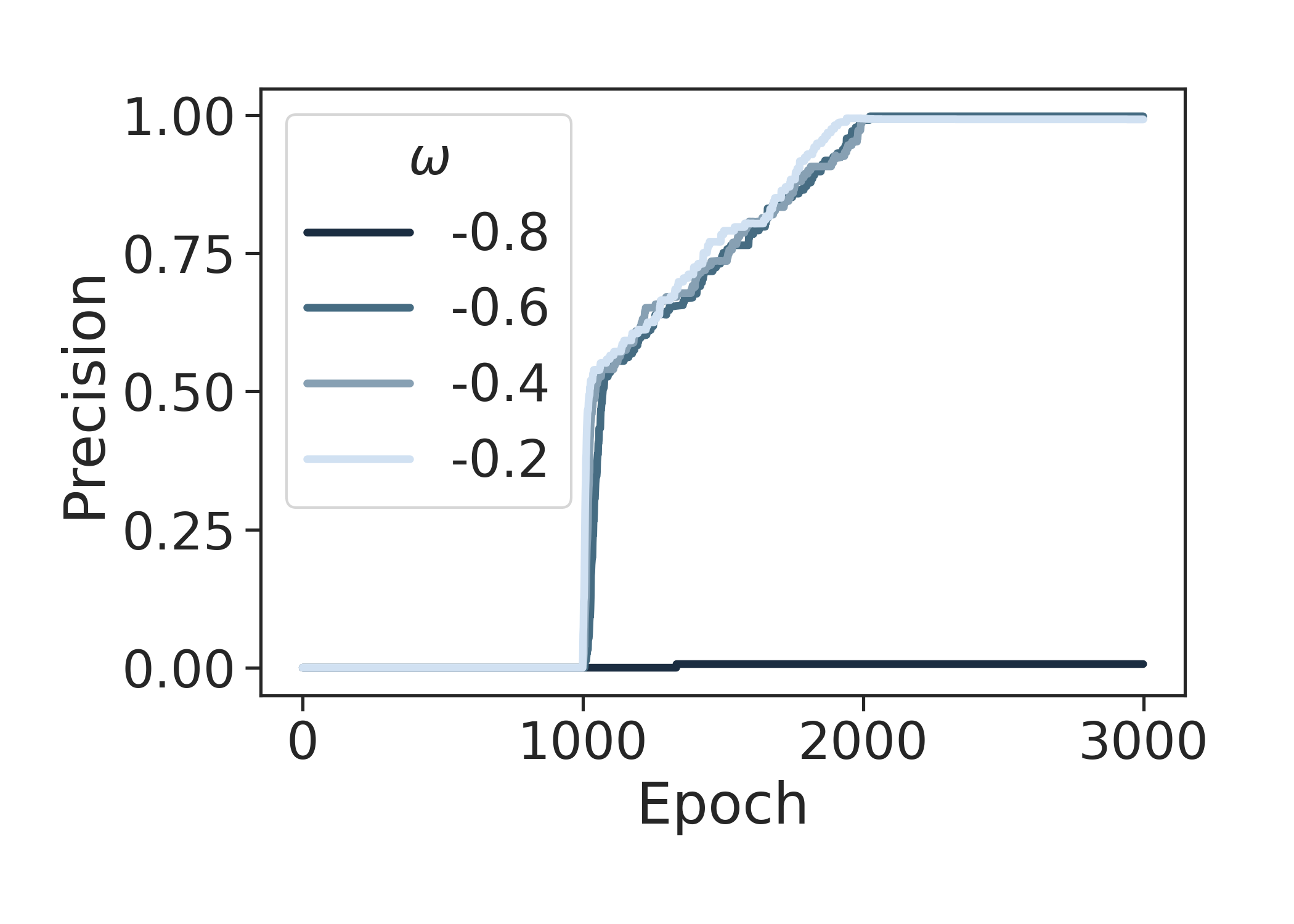}
            \label{fig:omega-prec}
        }
        \subfigure[Blacklisting Recall over $RMI$]{
            \includegraphics[width=0.23\textwidth]{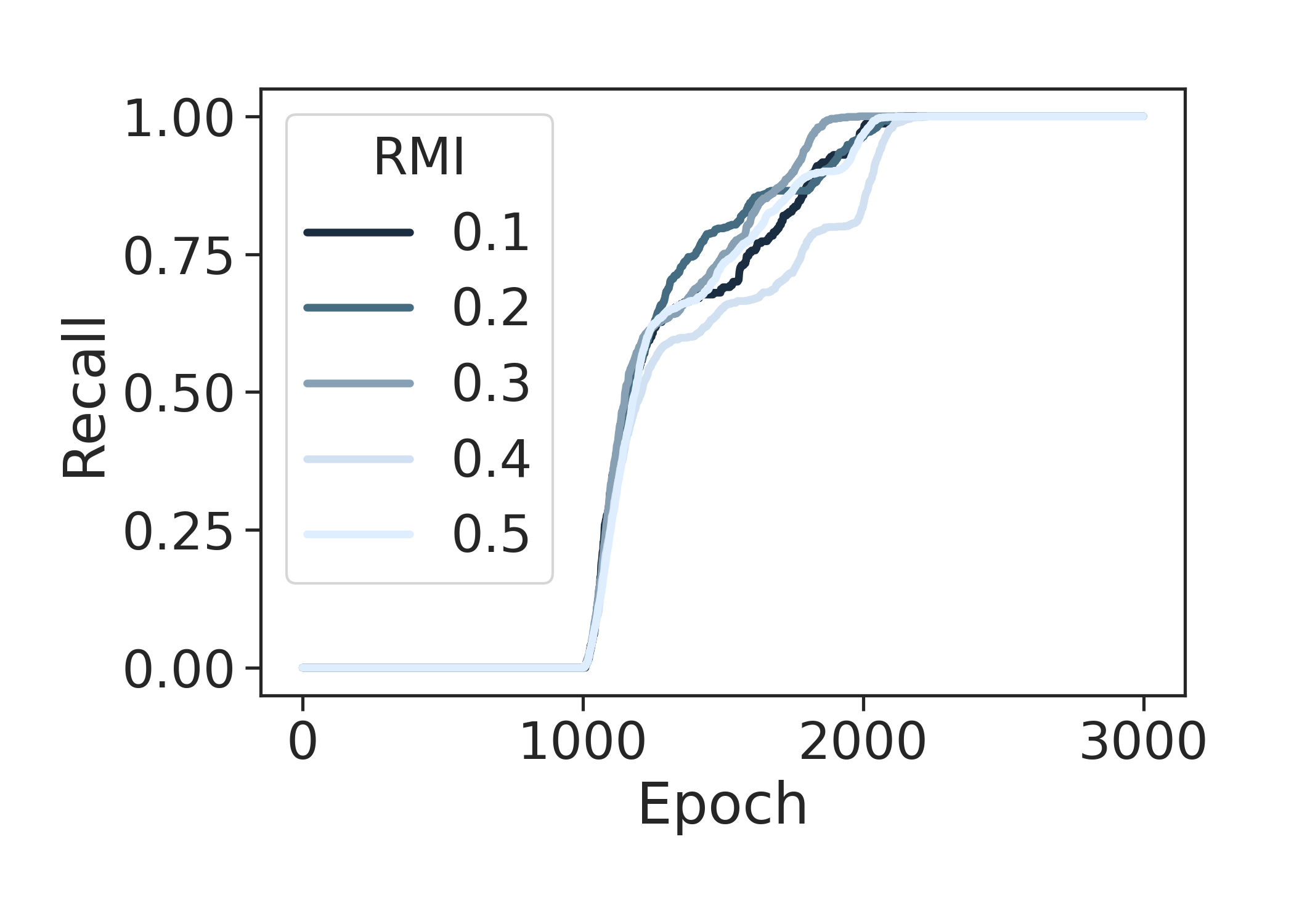}
            \label{fig:rmi-rec}
        }
         \subfigure[Blacklisting Precision over $RMI$]{
            \includegraphics[width=0.23\textwidth]{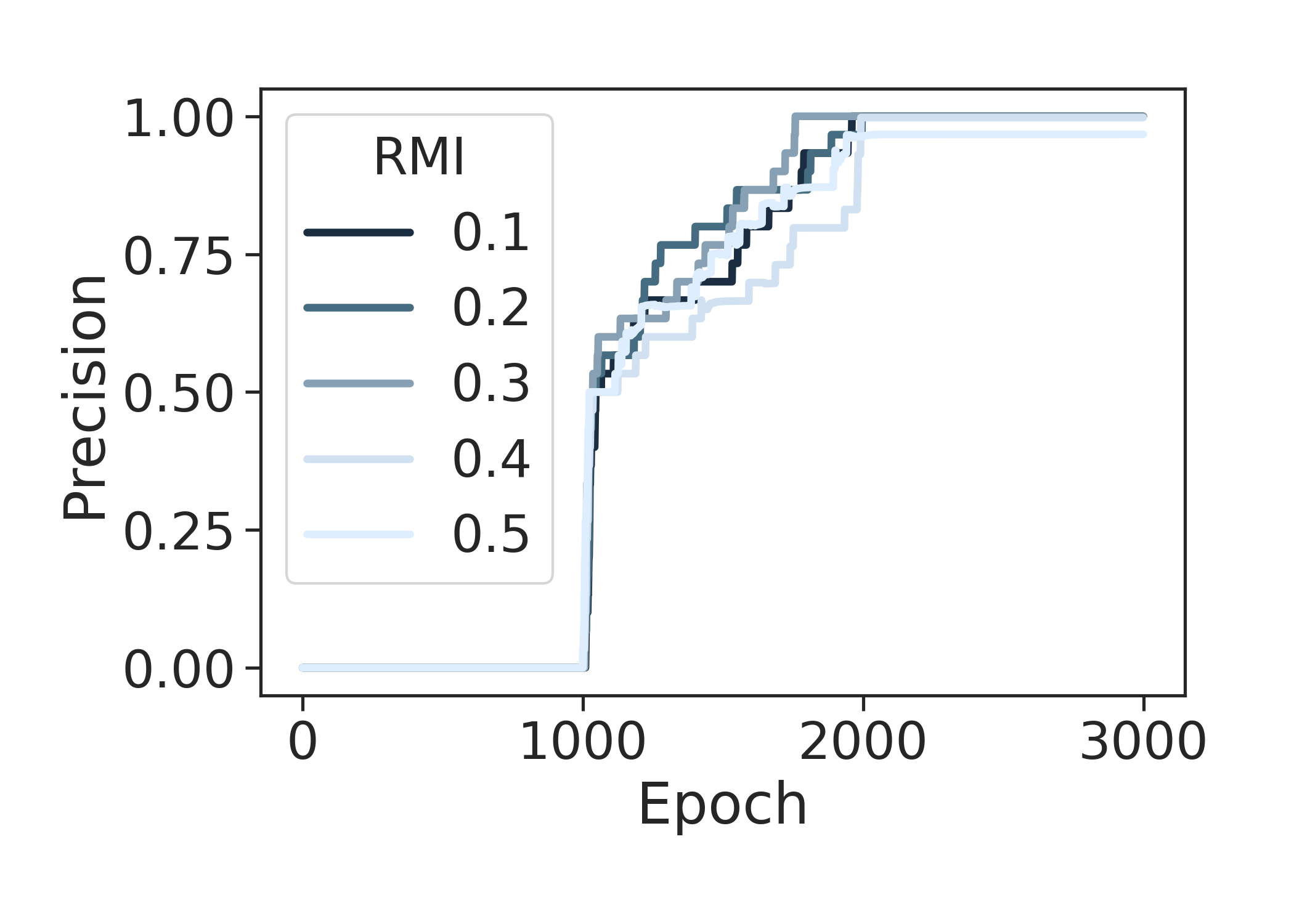}
            \label{fig:rmi-prec}
        }

    \caption{Behavior of the blacklisting action in Med-RB.}
    \label{fig:blacklisting_omega}
\end{figure*}

\subsection{Results}\label{sub:results-sec}

Our first experiment aims to validate the solidity of our proposed system against the collusion of a number of malicious Indexers. 
In this first batch of experiments we varied $RMI = \{0.1,0.2,0.3,0.4,0.5\}$ and $\omega = \{-0.8, -0.6, -0.4, -0.2\}$. All experiments were performed against all three methods (Med, Med-R, and Med-RB), and for each of them, we alternatively set up the malicious Producer arrival as either \textit{uniform} or \textit{bursty}. For each configuration, we performed 15 repetitions. This first experiment studies the impact of our system on malicious Indexers, thus we ran it with all Oracles being honest.

Fig.~\ref{fig:accuracy} shows the results in terms of truth inference accuracy, i.e., how many requests receive a response that does not deviate by more than $\tau_S$ from the ground truth. This number is calculated only during the last 1,000 epochs, where all malicious sources have made their appearance and have been around for a while and the system is thus in its steady state. Results show how the reputation mechanism has, in general, a very positive impact on the accuracy. The accuracy of Med constantly drops as malicious Indexers increase, while Med-R and Med-RB show a significant resilience that yields almost perfect results even with 40\% of Indexers being malicious. We expect the system to show similar, if not better, performance against defective or low-quality Producers, which represent still a better scenario than a joint attack of malicious ones. When malicious Indexers are around 50\% it is clearer how a bursty arrival has a negative impact versus a uniform one, however, our system still shows solidity, especially when performing blacklisting. 

At this stage, the following question arises: if blacklisting is too severe, are we risking to cut out potential honest Indexers in favor of reaching a high accuracy?

We then restricted our analysis to Med-RB to see whether this is happening, Fig.~\ref{fig:blacklisting_RMI} shows the analysis by varying the $RMI$, recall that here we are still varying $\omega$ as stated above, even though these plots show aggregate results.
Fig.~\ref{fig:rmi-rep}, in particular, shows the average reputation of honest and malicious Indexers separately over the course of the simulation.
Solid lines account for honest Indexers and dashed lines for malicious ones. 
No matter the value of $RMI$ (which here produces lines of different colours), the reputation of honest Indexers quickly adjusts to 0.6, while, symmetrically, the reputation of malicious Indexers settles to -0.6.
Furthermore, Fig.~\ref{fig:rmi-rec} shows the blacklisting recall over the course of the simulation, which means how many of all malicious Indexers get blacklisted. Once again, no matter the value of $RMI$, by the end of the simulation Med-RB manages to blacklist all malicious Indexers.
A similar behavior can be observed in Fig.~\ref{fig:rmi-prec}, which shows the blacklisting precision, i.e., of all blacklisted Indexers how many are actually malicious. In this case, we can see that the blacklisting strategy makes a handful of mistakes (less than 5\%) in the long run for $RMI=0.5$, while in all other cases, the average is set to 1, with minor fluctuation. This explains how, in the presence of many Indexers, Med-RB tends to kick out of the system very few honest (but less efficient) Indexers to maintain a high resilience to mistakes.

We can deepen our analysis by looking at the blacklisting precision and recall metrics over $\omega$ reported in Fig.~\ref{fig:blacklisting_omega}. In both plots, we can observe that $\omega=-0.8$ almost always fails to blacklist anything, as it is too permissive for our setup. Conversely, all other values settle to 1 for both precision and recall in the long run, with a slight exception of the blacklisting precision when $\omega = -0.2$, a configuration at times too severe that can occasionally kick out honest Indexers. Even though this happens in very few cases, we observe that $\omega =-0.4$ and $\omega =-0.6$ yield almost perfect results.

\begin{figure*}[ht!]
    \centering
        \subfigure[Truth Inference Accuracy for Med]{
            \includegraphics[width=0.31\linewidth]{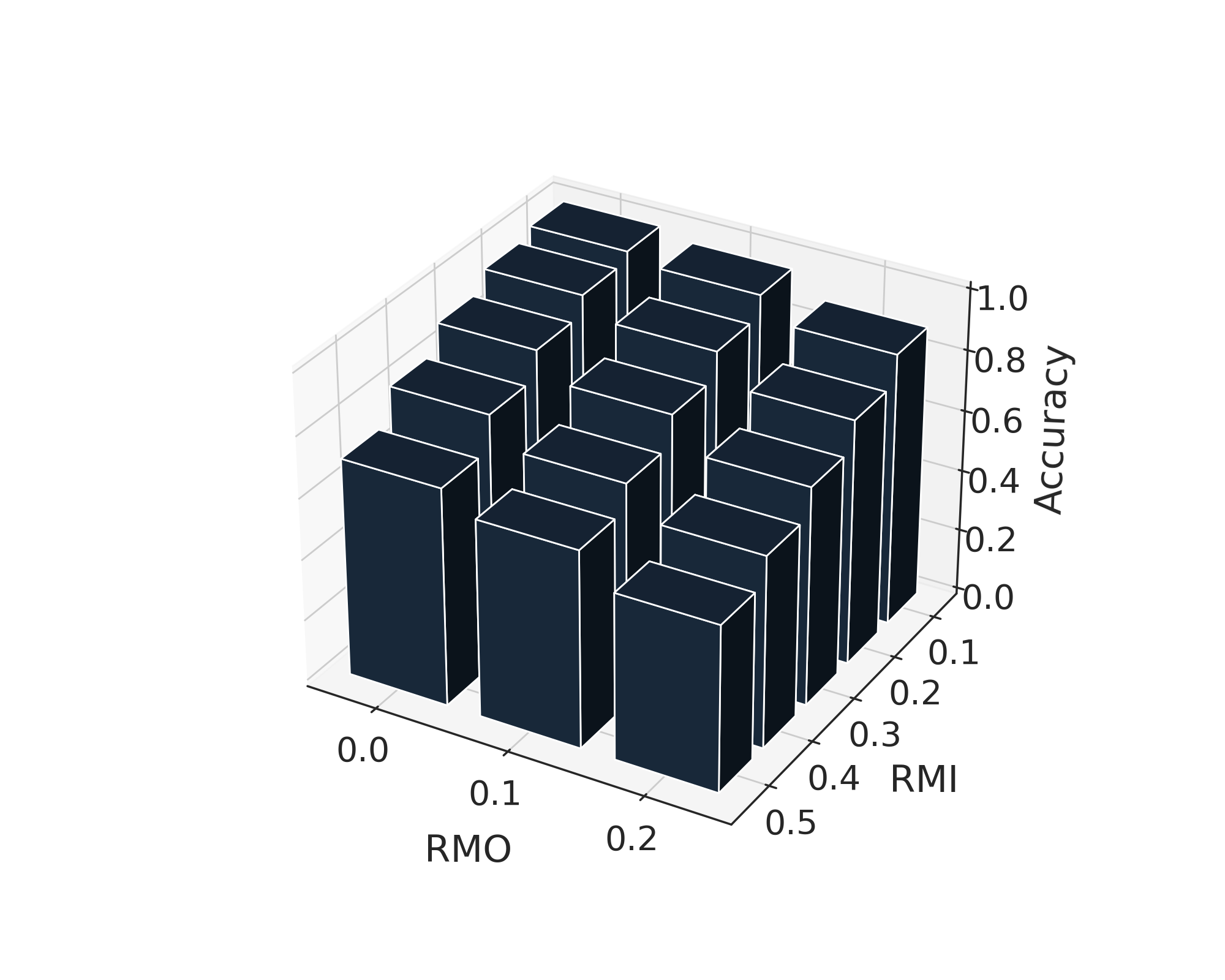}
            \label{fig:3d-med}
        }
         \subfigure[Truth Inference Accuracy for Med-R]{
            \includegraphics[width=0.31\linewidth]{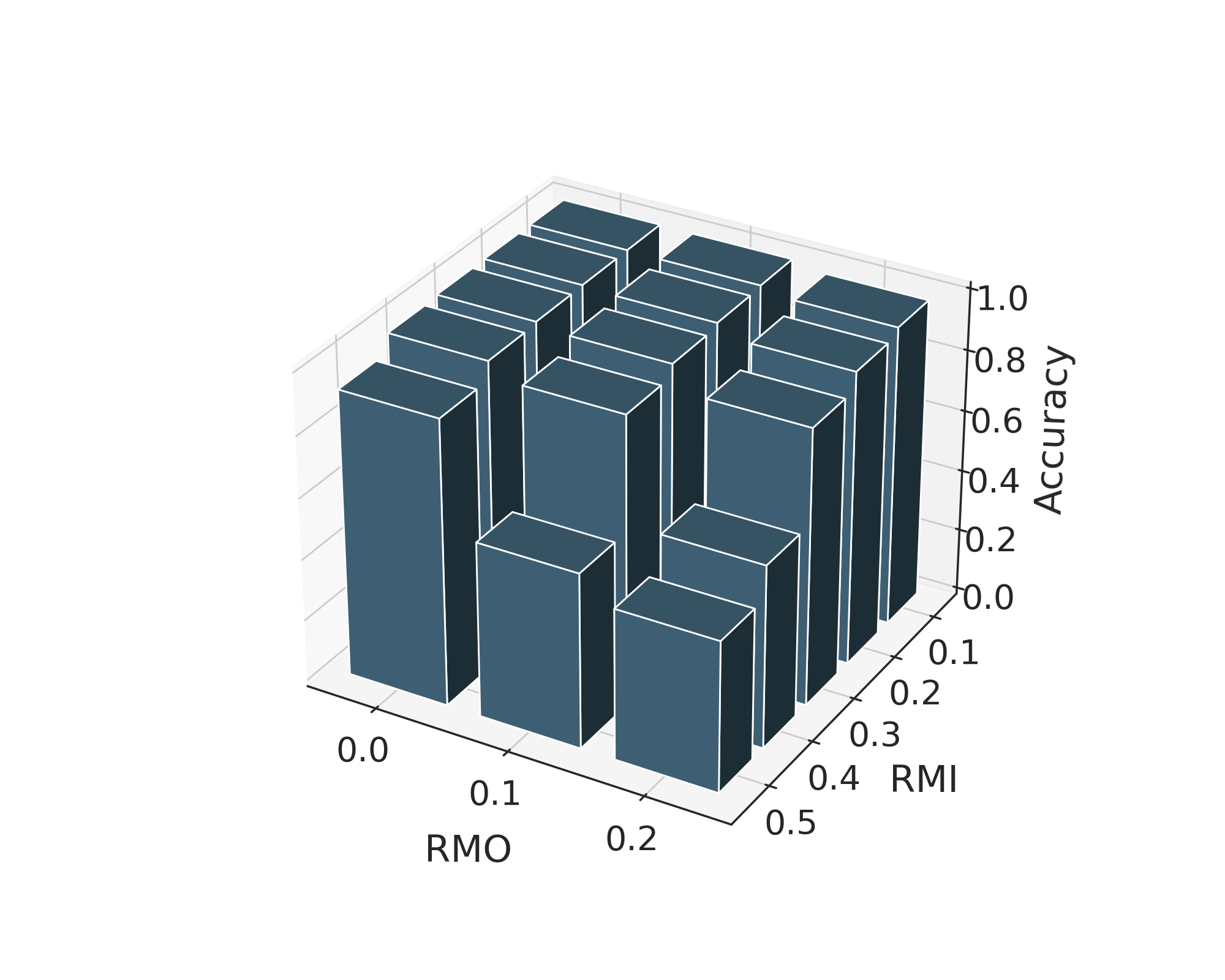}
            \label{fig:3d-medr}
        }
         \subfigure[Truth Inference Accuracy for Med-RB]{
            \includegraphics[width=0.31\linewidth]{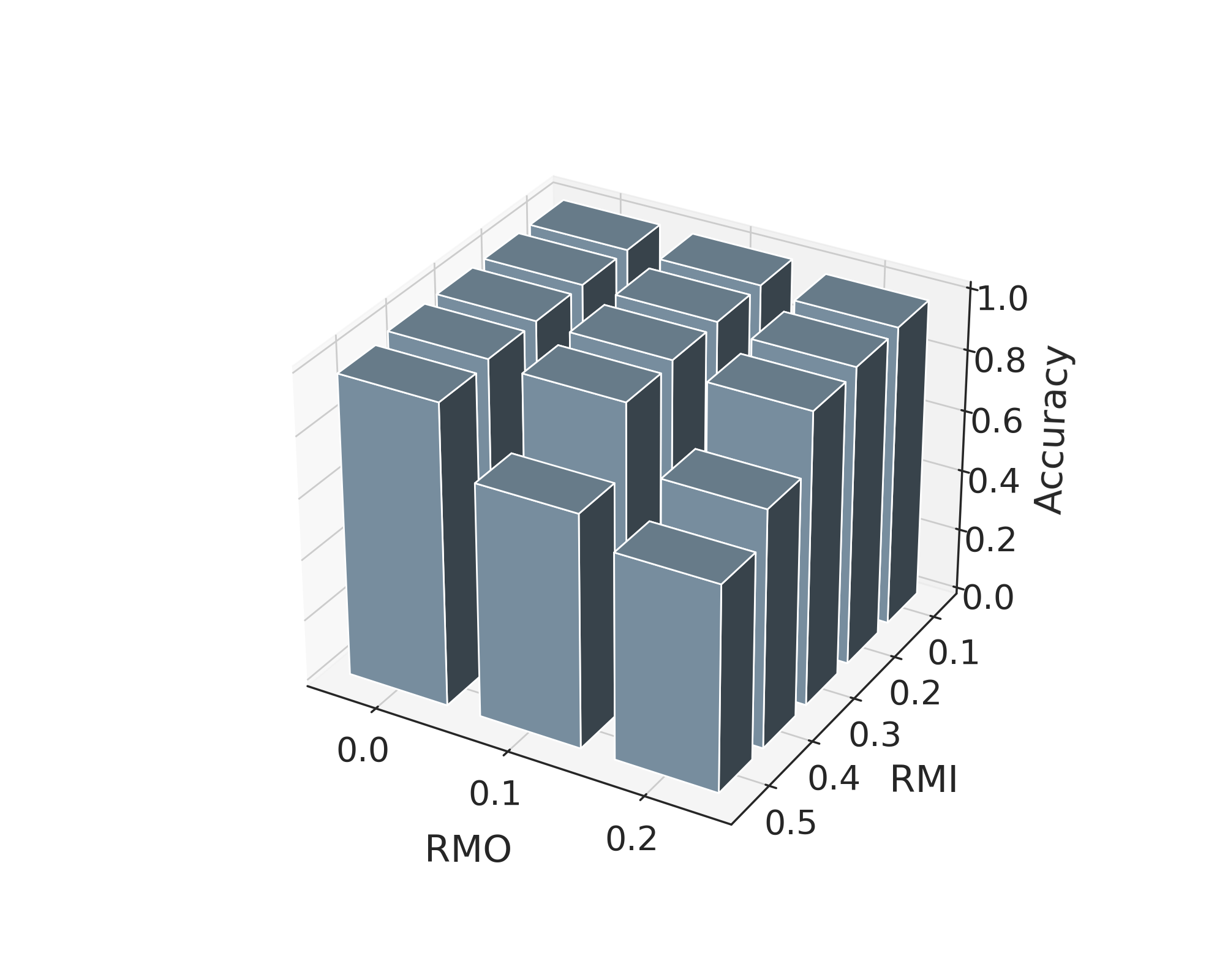}
            \label{fig:3d-medrb}
        }
    
    \caption{3D bar charts showing the Truth Inference Accuracy by varying both $RMI$ and $RMO$ for all three considered methods.}
    \label{fig:blacklisting_RMI}
\end{figure*}

In a second experiment, we run the same simulations, but varying also $RMO$ such that $RMO = \{0.0,0.1,0.2\}$. The results are shown by means of 3D bar charts in Fig.~\ref{fig:blacklisting_RMI}.
From the plots, it is evident how both methods based on reputation can hold good resilience to malicious entities, but when the collusion of malicious Oracles and Indexers start to take over, then the performance drops abruptly. In fact, if we consider the data matrix $\mathcal{D}$ and envision a situation where $RMI$ and $RMO$ also apply to the selected Oracles and Indexers for that round, we can estimate the ratio of elements in the matrix corrupted by either a malicious Oracle or Indexer as $RMO + RMI - (RMO\cdot RMI)$. If this number is greater than 0.5, then we can expect, on average the malicious entities to take over and invert the behavior of the system. In our simulations, this may be slightly mitigated by the warm start and this value may be higher than 0.5, however, we experimentally find that, for example, for $RMI=0.4$ and $RMO=0.2$ we estimate only 52\% of $\mathcal{D}$ to be corrupted, however, this is sufficient to make the performance drop.

\section{Related Work}\label{sec:related-work}

\begin{table*}[b]
\label{tab:related_work}
\caption{Comparison of existing decentralized oracle solutions upon ZONIA architectural features}
\begin{tabular}{|c|c|c|c|c|c|}
\hline
\textbf{Paper}                        & \textbf{Zero-Trust Design} & \textbf{Semantic and Geospatial Requests} & \textbf{Fair and Secure Node selections} & \textbf{Hidden Data Sources} & \textbf{Reputation} \\ \hline
\cite{chen2021tora}          & \ding{55}                                        & \ding{55}/\ding{55}                                       & \ding{51}/\ding{51}                                     & -                            & \ding{51}                   \\ \hline
\cite{woo2020distributed}    & \ding{55}                                        & \ding{55}/\ding{55}                                       & \ding{55}/\ding{51}                                      & -                            & \ding{51}                   \\ \hline
\cite{moudoud2021towards}    & \ding{51}                                        & \ding{55}/\ding{55}                                       & \ding{55}/\ding{51}                                      & -                            & \ding{51}                   \\ \hline
\cite{adler2018astraea}      & \ding{51}                                        & \ding{55}/\ding{55}                                       & \ding{55}/\ding{55}                                      & \ding{55}                            & \ding{55}                   \\ \hline
\cite{xian2024distributed}   & \ding{55}                                        & \ding{55}/\ding{55}                                       & \ding{55}/\ding{51}                                      & -                            & \ding{51}                   \\ \hline
\cite{moudoud2019iot}        & \ding{51}                                        & \ding{55}/\ding{55}                                       & -/-                                      & -                            & \ding{55}                   \\ \hline
\cite{du2022novel}           & \ding{51}                                        & \ding{55}/\ding{55}                                       & \ding{51}/\ding{51}                                      & -                            & \ding{55}                   \\ \hline
\cite{peterson2015augur}     & \ding{51}                                        & \ding{55}/\ding{55}                                       & \ding{51}/\ding{51}                                      & \ding{55}                            & \ding{51}                   \\ \hline
\cite{nelaturu2020public}    & \ding{51}                                        & \ding{55}/\ding{55}                                       & \ding{51}/\ding{51}                                      & \ding{51}                            & \ding{55}                   \\ \hline
\cite{xiao2023decentralized} & \ding{51}                                        & \ding{55}/\ding{55}                                      & \ding{51}/\ding{51}                                      & \ding{51}                            & -                   \\ \hline
ZONIA                                 & \ding{51}                                        & \ding{51}/\ding{51}                                       & \ding{51}/\ding{51}                                      & \ding{51}                            & \ding{51}                   \\ \hline
\end{tabular}
\end{table*}

In this section, we review existing literature related to oracles in blockchain technology, focusing on their design, security considerations, architectures, and applications. First oracle proposals~\cite{provable}~\cite{zhang2016town}~\cite{park2020smart} were based on centralized strategies, where a single entity is in charge of retrieving data to be provided to a smart contract. Since this approach has the main drawback of a single point of failure, more recent works adopt decentralized solutions that operate through multiple nodes that provide data to the blockchain, where each node typically depends on one or more designated data sources to fulfill requests. This is the case of~\cite{chen2021tora}, where authors present a trusted blockchain oracle based on a decentralized TEE layer-2 network called \textit{Tora}, with a focus both on efficiency and availability. The idea is to tackle the availability problem of centralized solutions by replicating the Oracle nodes, introducing a hybrid consensus mechanism with the available nodes selection based on a Proof-of-Availability (PoA) metric. More in detail, the hybrid consensus is reached among off-chain nodes run in TEE enclave, while the PoA measurement of each node depends both on the number of requests received and the number of requests successfully, with an incentive mechanism to maintain overall high availability across the network. In order to tackle the problem of low-quality or compromised external data, authors of~\cite{xiao2023decentralized} propose \textit{DECENTRUTH}, a decentralized truth-discovering oracle architecture to address the truthful data challenge using a data-centric approach. In particular, the solution presented harmonizes a novel composite batch incremental Truth Discovery (\textit{CBI-TD}) process with a consensus protocol (\textit{WP-ACS}) that enables nodes to propose its local truth estimates and jointly decide on a common subset of proposals under asynchronous network conditions. Priority is given to the proposals from nodes with higher historical weights computed by the CBI-TD. 

There exist approaches that, unlike traditional decentralized oracles, rely on human intelligence instead of automatic mechanisms for data certification. \textit{ASTRAEA}~\cite{adler2018astraea} is a decentralized oracle system based on a voting game involving two types of participants: voters and certifiers.  Here, a submitter selects propositions to be verified by allocating funds. Voters then randomly select and vote on available propositions. Certifiers, on the other hand, choose propositions and place substantial deposits to certify them as true or false. The certification outcome is determined by the sum of certifications weighted by the deposits. A similar approach is presented in~\cite{nelaturu2020public}, where a decentralized oracle system based on a voting-based mechanism to determine the truth or falsity of binary queries is introduced. Users submit queries as True, False, or Unknown, with reporters (or certifiers) responding by staking monetary incentives. The authors present a formal analysis demonstrating that the system incentivizes truthful reporting, eventually achieving a Nash equilibrium.  Also~\textit{Augur}\cite{peterson2015augur} relies on human activity. This is a decentralized oracle and platform for prediction markets, operating on a trustless model. Users holding Augur's native Reputation token determine market outcomes by staking their tokens on observed results. They receive settlement fees as rewards, fostering a system where truthful reporting is incentivized for Reputation token holders.

Finally, although most of the oracles are general-purpose and hence can retrieve data from IoT devices, specialized solutions for this task have been investigated in the literature. In~\cite{woo2020distributed}, authors leverage SGX and TLS for proposing DiOr-SGX, a distributed oracle for IoT data with the goal of minimizing the response time when an oracle node is malicious or overloaded, still granting data integrity. Also \textit{STB}~\cite{moudoud2021towards} is a scalable trustworthy blockchain architecture designed for the IoT, leveraging sharding and a Peer-to-Peer oracle network to ensure efficient, scalable, and trustworthy data exchange among unreliable IoT devices; the P2P oracle network verifies the data queries and authenticates its source, i.e., it ensures the truth value of IoT
retrieved data. This is achieved through a lightweight form of consensus, reached among the participants of the network by providing a verification probability and then computing the average. Similarly, \textit{LC4IoT}~\cite{moudoud2019iot} is a lightweight consensus proposed for a blockchain architecture with distributed oracles specifically designed for the supply chain. Every oracle votes the veracity of sensor data and the consensus is reached if the sum of positive votes exceeds a certain threshold. In~\cite{xian2024distributed} authors present a distributed and efficient oracle solution tailored for the IoT, enabling fast acquisition
of real-time off-chain data. The proposed architecture combines both TEE devices and ordinary devices to improve the system scalability and address the interoperability issue with the IoT devices. In order to enhance the efficiency of data collection, authors introduced a Quality of Service (QoS) metric that takes into account both response time and result accuracy of nodes retrieving data from the source. Then, they employ an efficient weighted random node selection algorithm to assign tasks to nodes, in a manner that significantly improves the system's service efficiency. Authors of~\cite{du2022novel} introduce oracle-aided IIoT blockchain (\textit{OIB}), a system designed to facilitate the implementation of smart contract-based IIoT applications. This framework leverages a distributed oracle network to provide seamless interaction between off-chain environments and smart contracts: the oracle system consists of both off-chain and on-chain components, enabling bidirectional communication between smart contracts. Specifically, the on-chain oracle contract can access information from non-oracle contracts and transmit off-chain data from oracles to non-oracle contracts. 

Table~\ref{tab:related_work} compares our proposal with the solutions presented in this Section: most of the works examined rely on a zero-trust design and provide a fair and secure node selection, while only a few of them explicitly mention the fact that there is a blind data source selection. Still, ZONIA is the only one offering semantic and geospatial requests by design with a reputation system for maintaining the reliability of the oracles network.

\section{Conclusions and Future Work}\label{sec:conclusions}
This work presented \archname, a blockchain oracle system tailored for IoT applications. \archname addresses critical data integrity, decentralization, and trust challenges, leveraging a zero-trust decentralized design that permits open participation without specialized requirements. The system supports semantic and geospatial requests, ensuring precise and context-aware data retrieval essential for various applications such as smart insurance and environmental monitoring processes. Node selection within \archname is fair and secure, employing a combination of randomness and reputation scores to prevent collusion and enhance system security.

Our evaluations demonstrate that \archname can scale, maintaining decentralization and supporting many malicious nodes without compromising reliability. The architecture ensures that data sources remain undisclosed to clients, enhancing data security and reducing manipulation risks. The comprehensive reputation system evaluates node reliability and contributes to a robust, tamper-resistant framework, ensuring data accuracy and trustworthiness.

For future work, we identified two significant areas of improvement. First, the integration of machine learning techniques for predictive analytics could be investigated to anticipate data reliability issues and dynamically adjust the network's trust parameters. Second, a distributed caching layer could be introduced to increase system performance, avoiding the entire resolution process for all the requests.

\bibliographystyle{ieeetr}

\begin{IEEEbiography}[{\includegraphics[width=1in,height=1.25in,clip,keepaspectratio]{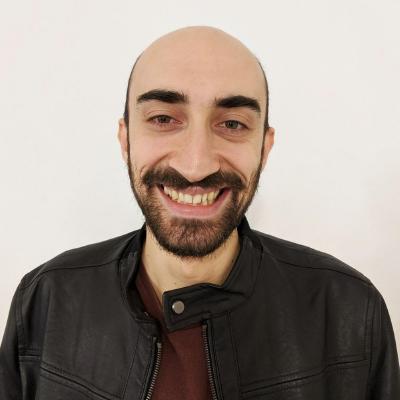}}]{Lorenzo Gigli} 
received his Master's Degree with distinction (summa cum laude) in Computer Science in 2019 from the University of Bologna, Italy. Subsequently, he served as a Research Fellow on the MAC4PRO project, funded by INAIL, at the Department of Computer Science and Engineering (DISI) at the same university. Currently, he is pursuing a Ph.D. in Engineering and Information Technology and is affiliated with the IoT PRISM laboratory, led by Prof. Marco Di Felice. His research interests encompass interoperability within the Internet of Things, distributed systems, and blockchain technologies.
\end{IEEEbiography}

\begin{IEEEbiography}[{\includegraphics[width=1in,height=1.25in,clip,keepaspectratio]{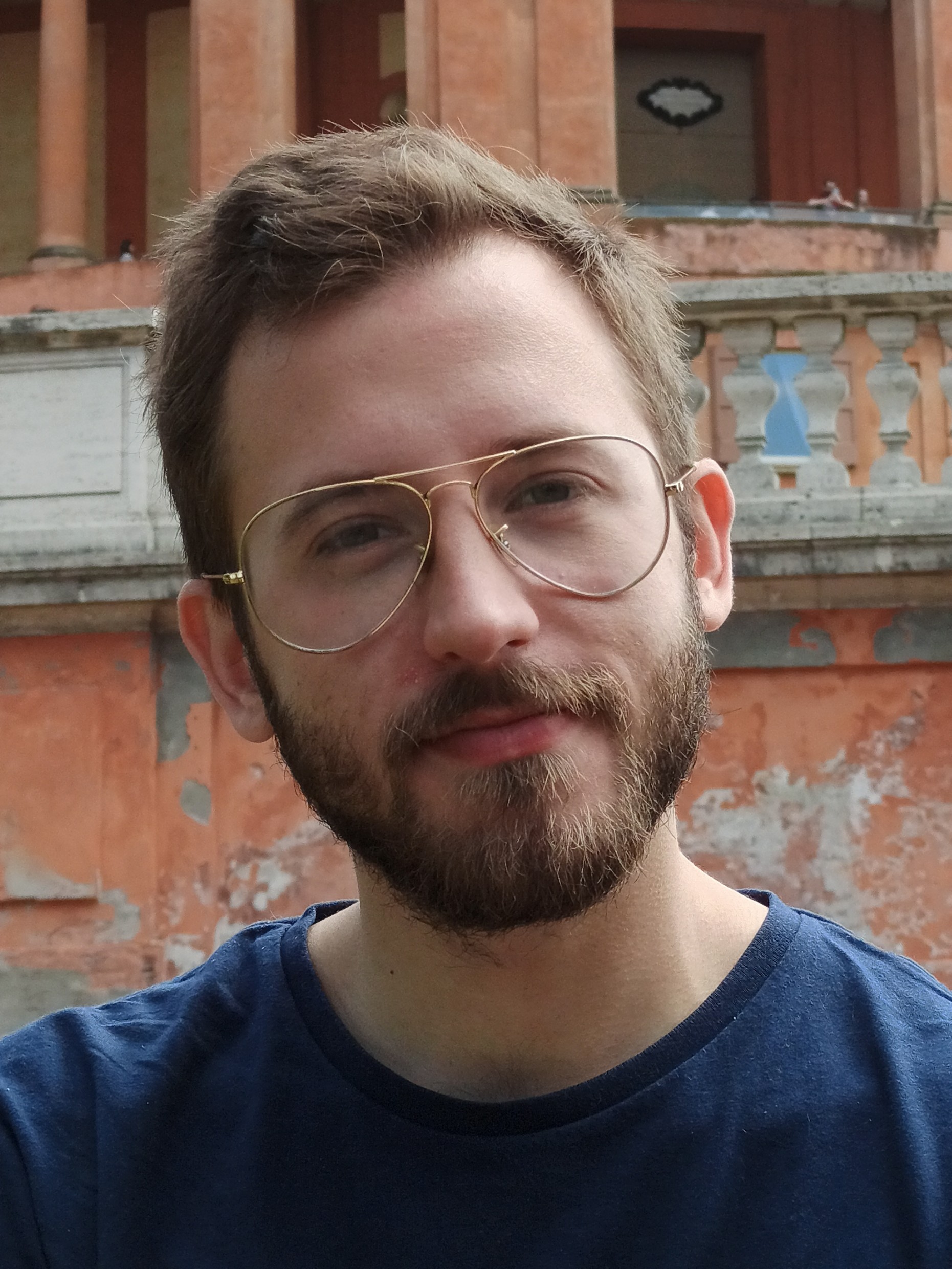}}]{Ivan Zyrianoff} is a Ph.D. student from the University of Bologna and a member of the IoT-Prism lab. He holds a B.S. degree in Computer Science and an M.S. degree in Information Engineering from the Federal University of ABC, Brazil. He was involved in the SWAMP Project, an EU-Brazil collaborative research project that developed IoT-based methods and approaches for smart water management in precision irrigation. He also participated in the Arrowhead Tools project, which aims for the digitalization and automation solutions for the European industry. His current research topics encompass interoperability for the Internet of Things and Edge Computing.
\end{IEEEbiography}

\begin{IEEEbiography}
    [{\includegraphics[width=1in,height=1.25in,clip,keepaspectratio]{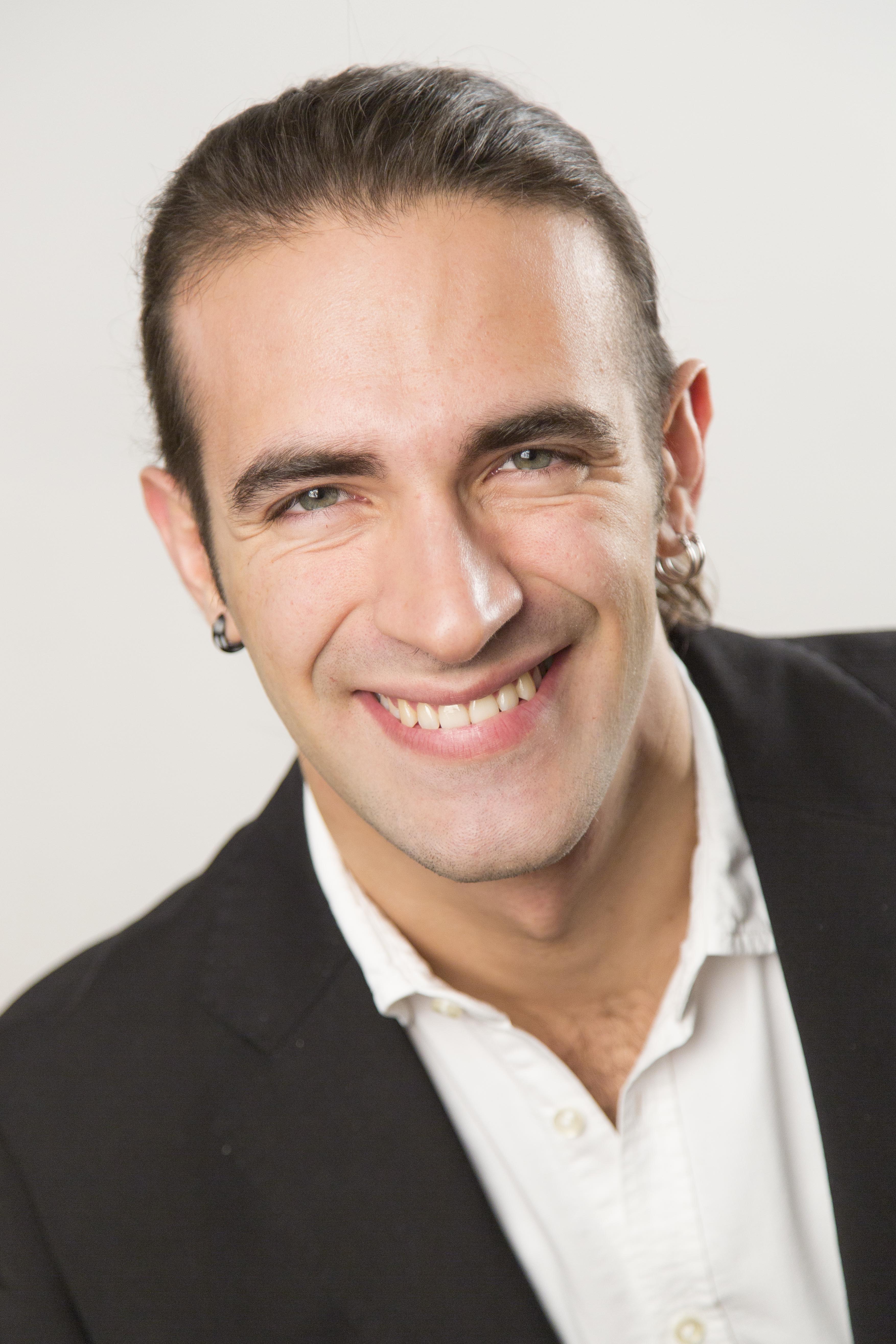}}]{Federico Montori} 
received the B.S. and M.S. degrees (summa cum laude) in computer science and the Ph.D. degree in computer science and engineering from the University of Bologna, Italy, in 2012, 2015, and 2019, respectively. He was a Visiting Researcher at Swinburne University of Technology (Australia), Luleå Tekniska Universitet (Sweden), and Technische Universit\"{a}t Ilmenau (Germany). He is currently a Senior Assistant Professor at the University of Bologna. He participated in several EU projects and was WP Leader for the H2020 Project Arrowhead Tools. His primary research interests include mobile crowdsensing (MCS), pervasive and mobile computing, IoT automation, and data analysis for IoT scenarios.
\end{IEEEbiography}

\begin{IEEEbiography}[{\includegraphics[width=1in,height=1.25in,clip,keepaspectratio]{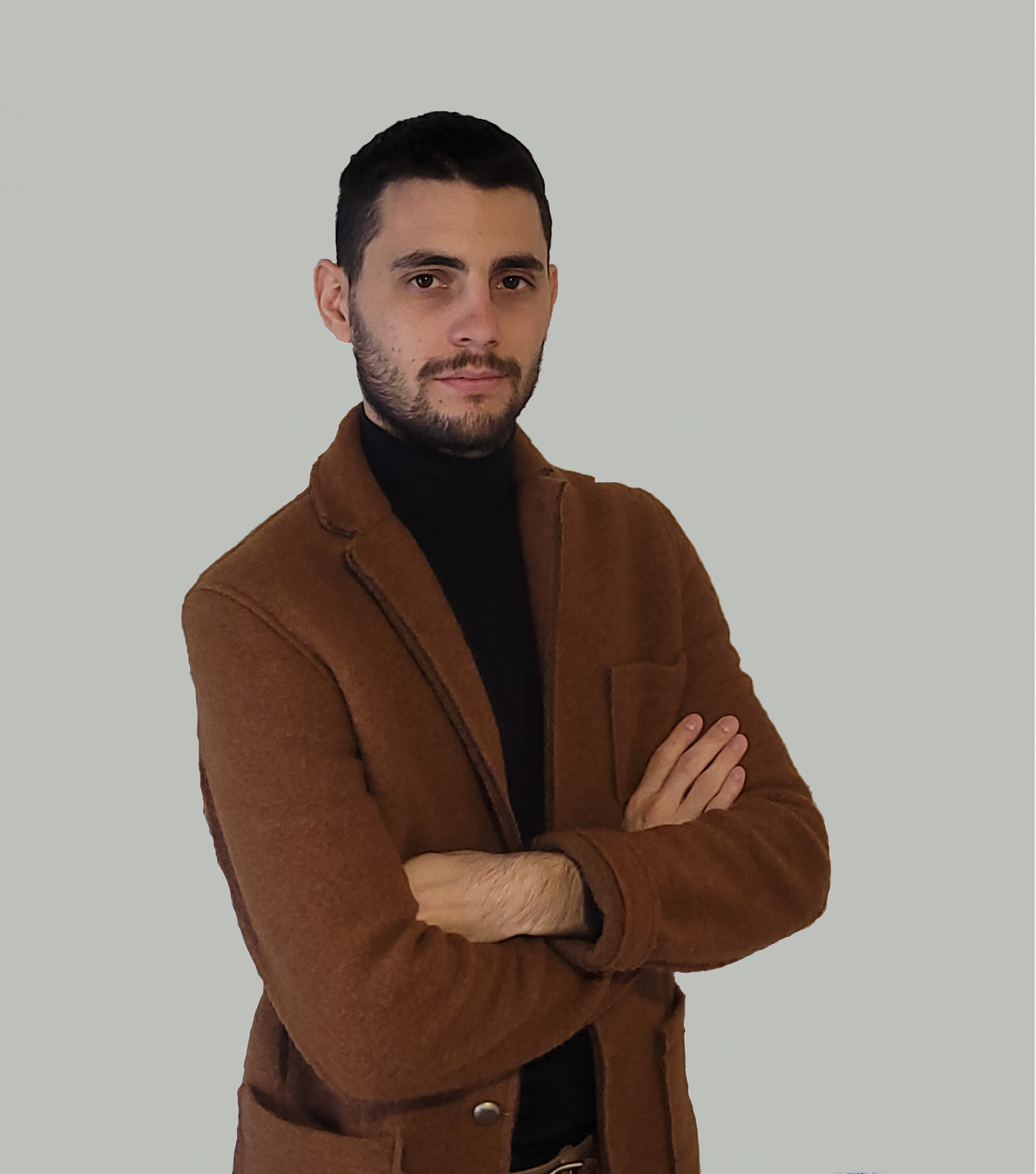}}]{Luca Sciullo} 
is a Junior Assistant Professor at the University of Bologna, Italy, where he also received the master’s degree (summa cum laude) in computer science and the Ph.D., respectively in 2017 and 2021. He is also a former Expert at the European Commission, where he explored the challenges related to ethics and privacy for the use of Personal Digital Twin (PDT). He was a Visiting Researcher at the Huawei European Research Center of Munich, Germany. He is a part of the IoT Prism Laboratory directed by Prof. Marco Di Felice and Prof. Luciano Bononi. His research interests include Data Interoperability, IoT systems, Web of Things, blockchain technologies, wireless sensor networks, Digital Twin, and Industry 4.0.
\end{IEEEbiography}

\begin{IEEEbiography}[{\includegraphics[width=1in,height=1.25in,clip,keepaspectratio]{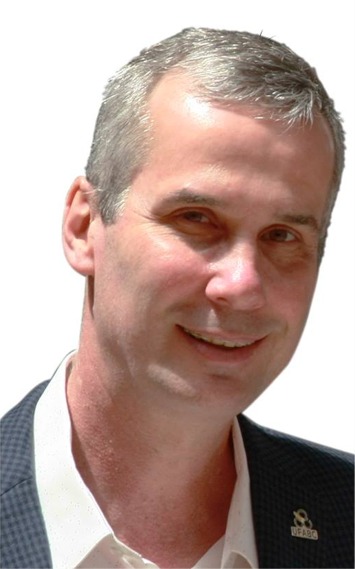}}]{Carlos A. Kamienski} is a Full Professor of Computer Science at the Federal University of ABC (UFABC, Brazil). For eight years, he led the NUVEM Strategic Research Unit comprising faculty members and students working in smart societies, virtual sensations, connected mobility, extreme computing, and integrated universes. He was the Brazilian coordinator of SWAMP from 2017 to 2021 (swamp-project.org), an EU-Brazil collaborative research project that developed IoT-based methods and approaches for smart water management in precision irrigation. His current research interests include the Internet of Things, smart agriculture, smart cities, fog computing, network softwarization, and Future Internet.
\end{IEEEbiography}

\begin{IEEEbiography}[{\includegraphics[width=1in,height=1.25in,clip,keepaspectratio]{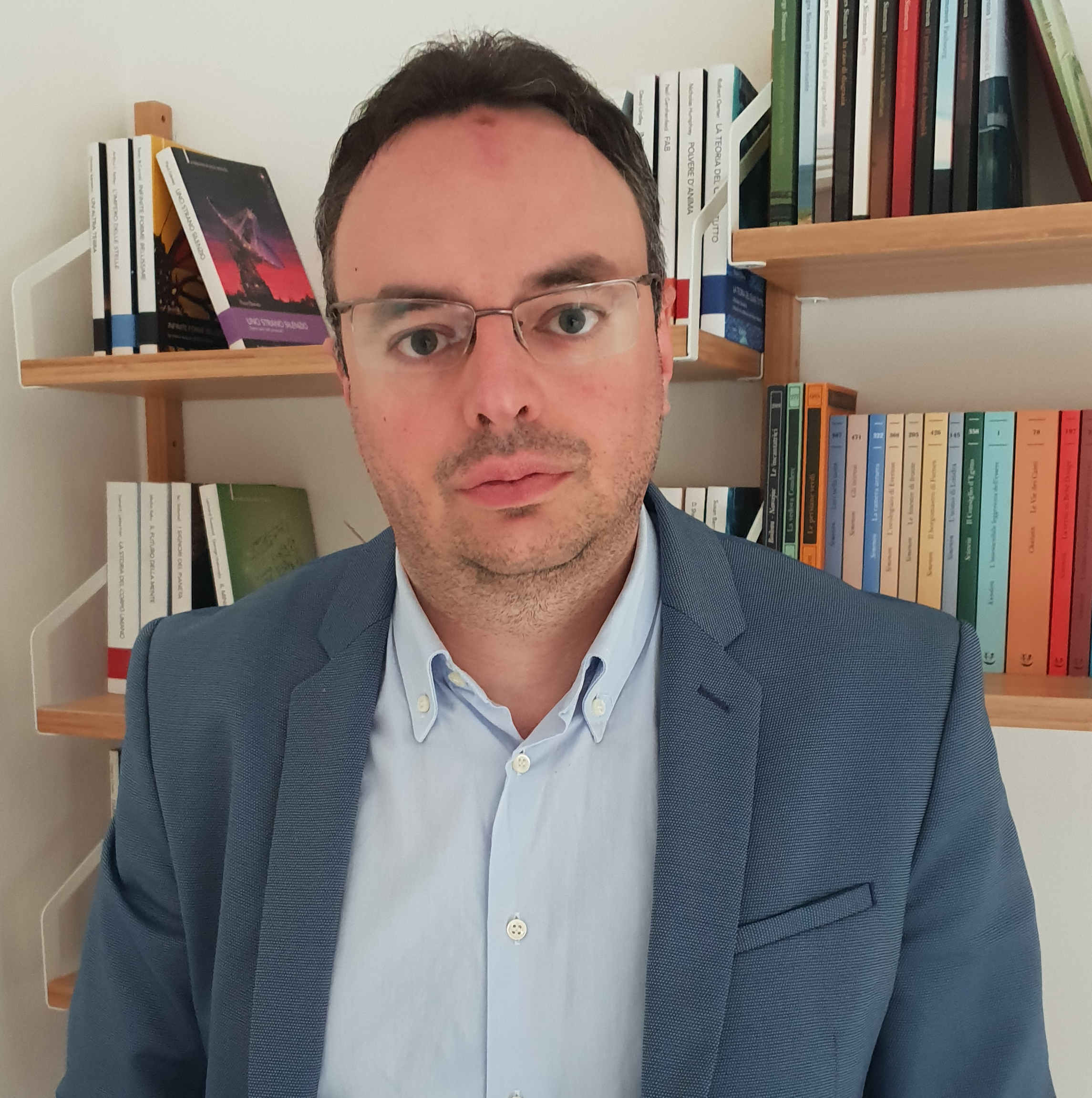}}]{Marco Di Felice} is a Full Professor of Computer Science with the University of Bologna, where he is the co-director of the IoT PRISM laboratory. He received the Laurea (summa cum laude) and Ph.D. degrees in computer science from the University of Bologna, in 2004 and  2008, respectively. He was a Visiting Researcher with the Georgia Institute of Technology, Atlanta, GA, USA and with Northeastern University, Boston, MA, USA. He authored more than 120  papers on wireless and mobile systems, achieving three best paper awards for his scientific production. He is Associate Editor of the IEEE Internet of Things journal.  His research interests include self-organizing wireless networks, unmanned aerial systems, IoT, WoT, and context-aware computing. 
\end{IEEEbiography}

\section*{Appendix}\label{appendix}

\subsection{Handling Null Values in Rating Functions}
In certain cases, Oracles may write a few null data points on the relay chain. In particular, a data point in $\mathcal{D}$ is null either when (i) the Producer did not respond within the timeout set by the querying Oracle (this may happen either because the Producer is faulty or because the Oracle's connection is too slow), (ii) the Producer responded with an invalid data point, either intentionally or involuntarily, or (iii) the Oracle willingly replaced the response of the Producer with a null value.
In any case, the two rating functions $\rho$ and $\theta$ are affected by null values, which can invalidate or distort the outcome of the rating process. For this reason, this section is dedicated to handling null values meaningfully to preserve the purpose of the rating functions.

First, if $\mathcal{D}_{i,j}$ is null, then we define $\mathcal{S}_{i,j} = \infty$, reason being the distance between a null value and the inferred ground truth must be as large as possible. This only affects the first member of Equation 6, since if $\mathcal{S}_{i,j} = \infty$, then $c(S(i,j), \tau_S ) = -1$ for any finite $\tau_S$. This results in punishing a Producer (and its Indexer) for outputting a null value regardless of who is to blame. The Producer can still obtain a fairly positive rating if its response to all other Oracles is good and consistent -- in other words if such a null value is an isolated case.

Second, if $\mathcal{D}_{i,j}$ is null, then we similarly define $\mathcal{V}_{i,j} = \infty$, provided that the null value is discarded when defining other elements of $\mathcal{V}_{*,j}$, as they are calculated using the column average.
This only affects the second member of Equation 6, since if $\mathcal{V}_{i,j} = \infty$, then $c(V(i,j), \tau_V ) = -1$ for any finite $\tau_V$. This similarly punishes a Producer proportionally to the number of null values it outputs (same as above), making this punishment independent from $\beta_\rho$. Differently, the second member of Equation 7 is negatively affected by $\mathcal{V}_{i,j} = \infty$, as $\sigma[V(*, j)]$ is also going to be $\infty$, leading to a dangerous corner case. 
Let us define $\varphi_i$ as \textit{the number of null values} in $\mathcal{D}_{i,*}$, with $0 \leq \varphi_i \leq K_{\mathcal{O}}$. 
Therefore, let us re-define Equation 7 as:
\begin{equation*}\label{eq:rateoraclesnull}
    \theta(i) = \beta_{\theta} c(t_i, \tau_t ) + (1 - \beta_{\theta}) \frac { \sum_{j=1}^{m} c'(\mathcal{D}_{i,j}) }{ m }
\end{equation*}
where
$$
c'(\mathcal{D}(i,j)) = \begin{cases}

    \frac{\varphi_i - K_{\mathcal{O}}}{K_{\mathcal{O}} - 1},& \text{if } \mathcal{D}_{i,j} \text{ is null}\\
    c(V(i,j), \sigma[V(*, j)] ),&  \text{otherwise}
\end{cases}
$$
With Equation above we ensure to punish also Oracles for retrieving null values from Producers, but only if this is an isolated case. In fact, it is easy to see that, if $\mathcal{D}_{i,j}$ is null, then the $O_i$ will receive $-1$ as a rating for such a data point only if $\varphi_i = 1$, that is, if $O_i$ was the only one retrieving a null value from the $j$-th Producer. Conversely, if the $j$-th Producer outputted a null value to all Oracles, that is, $\varphi_i = K_{\mathcal{O}}$, then all Oracles will receive a neutral rating of $0$ for such a data point, resembling the fact that most probably the Producer is faulty and Oracles are not to blame for retrieving a null value.

\end{document}